%% file: main.tex
\documentclass[conference,letterpaper]{IEEEtran}
\IEEEoverridecommandlockouts

\usepackage{hyperref}
\usepackage{amsmath}
\usepackage{xcolor}
\usepackage{colortbl}
\usepackage{listings}
\usepackage{tabularx}

\definecolor{codegreen}{rgb}{0,0.6,0}
\definecolor{codegray}{rgb}{0.5,0.5,0.5}
\definecolor{codepurple}{rgb}{0.58,0,0.82}
\definecolor{backcolour}{rgb}{0.95,0.95,0.92}

\lstdefinestyle{XMLStyle}{
  language=XML,
  keywordstyle=\color{blue},
  numberstyle=\tiny\color{codegray},
  stringstyle=\color{codepurple},
  morekeywords={E,S,M,T,Failure,line},
  basicstyle=\tiny\ttfamily,
  columns=fullflexible,
  keepspaces=true,
  frame=none,
  breaklines=true,
  aboveskip=0pt,
  belowskip=0pt
}

\usepackage[normalem]{ulem}
\usepackage{listings}
\input{macros.tex}

\begin{document}

\title{230,439 Test Failures Later: An Empirical Evaluation of Flaky Failure Classifiers\thanks{This work was supported in part by NSF grant number 2100037}}

\author{\IEEEauthorblockN{Abdulrahman Alshammari}
\IEEEauthorblockA{\textit{George Mason University}\\
Fairfax, United States \\
aalsha2@gmu.edu}
\and
\IEEEauthorblockN{Paul Ammann}
\IEEEauthorblockA{\textit{George Mason University}\\
Fairfax, United States \\
pammann@gmu.edu}
\and
\IEEEauthorblockN{Michael Hilton}
\IEEEauthorblockA{\textit{Carnegie Mellon University} \\
Pittsburgh, United States \\
mhilton@cmu.edu}
\and
\IEEEauthorblockN{Jonathan Bell}
\IEEEauthorblockA{\textit{Northeastern University} \\
Boston, United States \\
j.bell@northeastern.edu}}

\maketitle
\begin{abstract}
Flaky tests are tests that can non-deterministically pass or fail, even in the absence of code changes.
Despite being a source of false alarms, flaky tests often remain in test suites once they are detected, as they also may be relied upon to detect true failures.
Hence, a key open problem in flaky test research is: How to quickly determine if a test failed due to flakiness, or if it detected a bug?
The state-of-the-practice is for developers to re-run failing tests: if a test fails and then passes, it is flaky by definition; if the test persistently fails, it is likely a true failure. However, this approach can be both ineffective and inefficient.
An alternate approach that developers may already use for triaging test failures is failure de-duplication, which matches newly discovered test failures to previously witnessed flaky and true failures.
However, because flaky test failure symptoms might resemble those of true failures, there is a risk of missclassifying a true test failure as a flaky failure to be ignored. 
Using a dataset of 498 flaky tests from 22 open-source Java projects, we collect a large dataset of 230,439 failure messages (both flaky and not), allowing us to empirically investigate the efficacy of failure de-duplication. 
We find that for some projects, this approach is extremely effective (with 100\% specificity), while for other projects, the approach is entirely ineffective.
By analyzing the characteristics of these flaky and non-flaky failures, we provide useful guidance on how developers should rely on this approach.

\end{abstract}

\input{sections/FinalStyle/introduction}

\input{sections/FinalStyle/approaches}

\input{sections/FinalStyle/evaluation.tex}

\input{sections/FinalStyle/threats}

\input{sections/FinalStyle/discussion}

\input{sections/FinalStyle/relatedWork}
\input{sections/FinalStyle/conclusion}

\clearpage

\bibliographystyle{IEEEtran}
\bibliography{sample-base}

\end{document}

%% file: macros.tex
\usepackage{booktabs}
\usepackage{xspace}
\usepackage{enumitem}

\newcommand{\failure}{failure message and stacktraces\xspace}
\newcommand{\failures}{failure messages and stacktraces\xspace}

\newcommand{\syntax}{text-based matching\xspace}
\newcommand{\classifier}{Failure Log Classifier\xspace}
\newcommand{\tfidf}{{TF-IDF}\xspace}

\newcommand{\alluxio}{\emph{Alluxio-alluxio}\xspace}
\newcommand{\okhttp}{\emph{Square-okhttp}\xspace}
\newcommand{\hbase}{\emph{Apache-hbase}\xspace}
\newcommand{\ambari}{\emph{Apache-ambari}\xspace}

\newcommand{\httpcore}{\emph{Apache-httpcore}\xspace}
\newcommand{\websocket}{\emph{Tootallnate-java-websocket}\xspace}

\newcommand{\wildfly}{\emph{Wildfly-wildfly}\xspace}

\newcommand{\spring}{\emph{Spring-projects-spring-boot}\xspace}

\newcommand{\elastic}{\emph{elasticjob-elastic-job-lite}\xspace}
\newcommand{\orbit}{\emph{orbit-orbit}\xspace}
\newcommand{\exec}{\emph{apache-commons-exec}\xspace}
\newcommand{\Achilles}{\emph{doanduyhai-Achilles}\xspace}

\AtBeginDocument{%
  \providecommand\BibTeX{{%
    \normalfont B\kern-0.5em{\scshape i\kern-0.25em b}\kern-0.8em\TeX}}}

%% file: sections/FinalStyle/introduction.tex
\section{Introduction}
\label{sec:intro}

Ideally, when an automated test case fails during development, this failure indicates that a defect has been detected.
Tests that detect defects are good tests, because they can signal developers to pause working on new development, and to debug and fix the defect promptly.
However, some test cases may be ``flaky,'' and can non-deterministically pass or fail, even when repeatedly executed on the same version of the same code.
For example, studies have shown that tests can be flaky due to strict reliance asynchronous computations completing in some specific order, reliance on undocumented platform dependencies, among other sources of randomness~\cite{Parry21Survey}.
When a test fails \emph{due to flakiness}, developers may still need to pause their development activities to confirm that the test failure does not actually represent a true defect.

While a flaky test is one that \emph{can} fail due to some non-deterministic reason, the failures of flaky tests cannot be entirely ignored, as flaky tests can \emph{also} detect defects.
For example, Rahman and Rigby studied the Mozilla Firefox continuous integration system, finding that when developers ignored failures of flaky tests, there was a dramatic increase in the number of crashes reported by users~\cite{Rahman18Impact}.
Similarly, Haben et al. analyzed 9 months of test failures in the Google Chromium continuous integration system, finding that ignoring \emph{all} failures of flaky tests would have resulted in missing 76\% of the true regression faults~\cite{haben2023importance}.
Surveys of developers report that flaky tests waste developers time and are a moderate-to-severe problem for most developers~\cite{Gruber22Survey,Eck19Understanding,habchi2022qualitative}.

In order to reduce the burden of inspecting every (possibly flaky) test failure, the state-of-the-practice approach for triaging test failures as flaky or true failures is to rerun failed tests~\cite{Parry21Survey,Gruber22Survey,Eck19Understanding,habchi2022qualitative}.
If, on the same version of the system under test, the test first fails, and then later the test passes, then it is a flaky failure that can be ignored.
Unfortunately, this approach is not guaranteed to detect all flaky failures, since a flaky test may also persistently fail.
Lam et al. observed that roughly half of the flaky tests in their dataset would persistently fail when re-run in isolation from the rest of the test suite~\cite{Lam20Understanding}.
Bell et al. studied the efficacy of Apache Maven's built-in test rerunning feature, finding that it could only confirm 23\% of flaky test failures as flaky~\cite{Bell18Deflaker}.
Particularly when tests might need to be re-run \emph{many} times, this procedure is expensive and time consuming.
A report from Google indicates that 2-16\% of computing resources are regularly dedicated just to re-running flaky tests~\cite{Micco17State}.

Hence, an important problem for flaky test research is: given a test failure, how to determine if this specific failure can safely be ignored (since it is flaky), or more thoroughly debugged (as a true failure).
While a significant body of academic literature aims to determine which tests are flaky tests (i.e., could exhibit flaky failures)~\cite{Lam20Understanding,alshammari2021flakeflagger,bell2018deflaker,lam2019idflakies,Pontillo22Static,Verdecchia21Know,Parry23Empirically,Lam20Study,Gyori16NonDex}, the problem of distinguishing flaky failures from true failures is understudied.
This article presents a large-scale empirical study that characterizes the design space for flaky failure detection, and preliminary results of several baseline approaches.
From 22 open-source Java projects, we collect a dataset of 498 flaky tests, including 80,530 flaky failures and 147,613 true failures.
This broad dataset complements existing case studies of flaky failures in Google Chromium~\cite{haben2023importance} and SAP HANA~\cite{An23JustInTime} by highlighting the variability of flaky failure detection across different projects.

We evaluate the use-case where a developer has an existing history of test failures (both flaky and true failures) and is given the task of determining if a new failure is flaky or not.
We find that flaky test failures can be extremely repetitive --- when a test fails due to flakiness, it is likely to match other flaky failures from the same or other tests.
We apply approaches based on failure de-duplication~\cite{Podgurski03Automated,Jiang17WhatCauses}, text-based matching, and simple machine learning classifiers.
We find that, for some tests, these approaches can be extremely effective (with no false negatives or false positives), yet for other tests, these approaches are entirely ineffective.
By examining attributes of tests and failures, we provide insights for future research on generalized approaches for detecting flaky failures.

The primary contributions of this paper are:
\begin{itemize}
    \item \textbf{Evaluation}: A large-scale evaluation of failure de-duplication using \failures to determine if a failure is flaky or true failure. 
    \item \textbf{Dataset}: An extended dataset that contains both flaky and true failures of tests, constructed using a novel approach based on mutation analysis. 
\end{itemize}

%% file: sections/FinalStyle/approaches.tex
\section{Matching Failures Logs}
\label{sec:approaches}

Failure logs provide a detailed understanding of the origin of the failure.
Hence, developers typically debug logs to better understand the failure cause. 
In detecting test flakiness, a recent survey shows that some developers may manually debug failures logs to tell if a failure is flaky or not~\cite{habchi2022qualitative}. 
Developers can recognize a failure is flaky by examining the \failure as they could have encountered flaky failures with similar \failure~\cite{gradlePreventingFlaky}.

We examine the applicability of failure  de-duplication to help developers to determine if a new failure is flaky or not.
We evaluate three approaches for failure de-duplication: \syntax, which uses the text of failure logs to find the similarity of given two failures, the \classifier, which adopts machine learning to predict if a failure is flaky or not, and TF-IDF (the details appears later in subsection~\ref{sec:tfidf}), which uses information retrieval techniques to group failures.
We focus specifically on matching the output that is common to the test suites of all projects that we have studied: stack traces.
When matching stack traces, we consider two use-cases: matching different failures from the same test, or matching failures from different tests.

\subsection{Text-Based Matching}
\label{syntax}
Text-based matching is our application of classic failure de-duplication approaches~\cite{Podgurski03Automated,Jiang17WhatCauses}, where we de-duplicate failures by matching common stack traces. 
This approach is also motivated by grey-literature suggestions that, ``sometimes it's obvious to engineers that a test is flaky just by looking at the exception type and message'' \cite{gradlePreventingFlaky}.
Intuitively, if an engineer has repeatedly seen the same flaky failure symptoms, they may be able to guess that a new failure is also flaky.
\syntax acts as an automated standin for this experience-based process.

We implement text-based matching by creating a dataset of parsed failure logs for each test.
Each failure log is represented by its \failure.
In terms of a failure message, it consists  \emph{exception type} (for example, AssertionFailedError) and everything follows this is treated as the \emph{exception message}. For the stack traces part, it is a set of lines representing the calls before the exception occurs and during the parsing, we are considering the top lines pointing directly to the test name. These lines reflect the most recent operations preceding the exception and often provide more details about the root cause of the failure.

We implement a pipeline to parse each failure into an XML file, cataloging all failures linked to a specific test.
As shown in Listing~\ref{lst:flakyFailures}, each failure block in the XML corresponds to one failure, containing four key components: the test name (\textbf{T}), exception type (\textbf{E}), exception message (\textbf{M}), and stacktrace lines (\textbf{S}). Within the \textbf{S} tag, individual lines are listed under the \textbf{line} tags, considering their order in the original log.
If the test name is missing from the stacktrace (e.g. it fails in setup method), we consider the last line from the test class. For example, in Listing~\ref{lst:flakyFailures}, the last line is not starting with the test name (present in \textbf{T}) but starts with the test class name. To categorize these XML files per project, the \textbf{T} tag includes a \emph{project} attribute, referring the project name where the test belongs.
In this phase, we also filter out non-deterministic stack trace lines internal to the JVM (e.g. \texttt{GeneratedMethodAccessor\$XYZ} lines).

\input{sections/Listings/FlakyFailures1}

The \syntax relies on the text of the \failure. As shown in Listing~\ref{lst:flakyFailures}, we found that the failure message (\textbf{M}) could contain information such as timestamp and IP address that make each (otherwise equivalent) failure unique.
For example, in Listing~\ref{lst:flakyFailures}, different details like an IP address within \textbf{M} can set two failures apart.
Hence, the \syntax does not rely on \textbf{M}, and consider only stack traces (\textbf{S}) and exception type (\textbf{E}).
Given the challenges in capturing all potential cases where the failure message (\textbf{M}) could be identical, we avoid modifying these unique message details and discard the \textbf{M} during the comparison. 

When given flaky and true failures, the \syntax should be able to tell if a new failure is a de-duplication of flaky failures, true failures, or both. 
As this approach is designed to find failure de-duplication within the same test, it could be useful to applicable across different tests especially if the failure stack traces do not cover the test body, similar to the example provided in Listing~\ref{lst:flakyFailures}.

\input{sections/Tables/Features}

\subsection{Failure Log Classifier}
\label{classifier}

There are cases where a newly written test introduces flakiness, or when there is no prior failures for reference.
Motivated by these scenarios, we propose the \classifier, which is trained on both flaky and true failures from \emph{all} tests in a test suite. Then the classifier would be able to predict if the new encounter failure is flaky or true failure.
For training the \classifier, we considered selected the features shown in Table~\ref{table:Features}, based on their generality. 
We chose the features based on the text of the failure logs. 
Although other studies for predicting flaky failures use dynamic details~\cite{lampel2021life}, our goal is to determine if relying on the information in failure logs can effectively predict flaky failures.

We employ a simple \emph{Decision Tree} as the supervised learning algorithm~\cite{DT}. Based on the binary features used to train the classifier, decision tree provides a clear way to handle non-linear relationships. As a comparison, we also evaluate the applicability of a Na\"{i}ve Bayes classifier as well.

\subsection{TF-IDF}
\label{sec:tfidf}

Term Frequency-Inverse Document Frequency (\tfidf) is a commonly used numerical statistic that reflects how important a word is to a document in corpus~\cite{tfidf}.
\tfidf has two components: Term Frequency (\emph{TF}) which represents the frequency of a term (word) in a document and if a term appears frequently in a document, its \emph{TF} will be high. Second, Inverse Document Frequency (\emph{IDF}) which measures the significance of the term in the entire corpus and if a term appears in many documents, its \emph{IDF} value will be low, reflecting its lower importance. The \tfidf value of a term in a document is the product of its \emph{TF} and \emph{IDF} values. Equation~\ref{TF} and~\ref{IDF} show the computation of \emph{TF} and \emph{IDF}, respectively. 

\begin{equation}
\label{TF}
\text{TF (\textit{t})} = \frac{\text{Number of times term \textit{t} in a document}}{\text{Total number of terms in the document}}
\end{equation}

\begin{equation}
\label{IDF}
\text{IDF (\textit{t})} = \log(\frac{\text{Total number of documents}}{\text{Number of documents where \textit{t} in it}})
\end{equation}

In the context of studying failure logs, we refer \emph{document} to a \emph{failure} and the \textit{t} to the token we extract from each \failure. 
For each failure in the generated XML file used in the \syntax, we tokenize each line of each stacktrace (including the exception type) by split the words using the \emph{dot} as separator (and removing the symbols such parentheses). 
For example, the last line in Listing~\ref{lst:flakyFailures} will be converted to the following set of tokens \textit{(tachyon, JournalTest, before, JournalTest, java, 33)}.
As our goal was to evaluate the overall potential for this approach, we did not consider more advanced tokenization approaches~\cite{tfidf1}.

%% file: sections/Listings/FlakyFailures1.tex
\vspace{5px}

\begin{lstlisting}[style=XMLStyle, caption=Two flaky failures reported in Alluxio project after parsing their failure logs,label=lst:flakyFailures]

<Failure>
    <T project="alluxio">tachyon.JournalTest.TableTest</T>
    <E>UnknownHostException</E>
    <M>ip-172-31-48-81: ip-172-31-48-81: Temporary failure in name resolution</M>
    <S><line>java.net.Inet6AddressImpl.lookupAllHostAddr(Native Method)</line>
    <line>java.net.InetAddress$2.lookupAllHostAddr(InetAddress.java:929)</line>
    <line>java.net.InetAddress.getAddressesFromNameService(InetAddress.java:1324)</line>
    <line>java.net.InetAddress.getLocalHost(InetAddress.java:1501)</line>
    <line>tachyon.LocalTachyonCluster.start(LocalTachyonCluster.java:104)</line>
    <line>tachyon.JournalTest.before(JournalTest.java:33</line></S>
</Failure>
    ...
<Failure>
    <T project="alluxio">tachyon.JournalTest.TableTest</T>
    <E>UnknownHostException</E>
    <M>ip-172-31-58-81: ip-172-31-58-81: Temporary failure in name resolution</M>
    <S><line>java.net.Inet6AddressImpl.lookupAllHostAddr(Native Method)</line>             
    <line>java.net.InetAddress$2.lookupAllHostAddr(InetAddress.java:929)</line>            
    <line>java.net.InetAddress.getAddressesFromNameService(InetAddress.java:1324)</line>            
    <line>java.net.InetAddress.getLocalHost(InetAddress.java:1501)</line>
    <line>tachyon.LocalTachyonCluster.start(LocalTachyonCluster.java:104)</line>
    <line>tachyon.JournalTest.before(JournalTest.java:33)</line></S>
</Failure>
\end{lstlisting}

\vspace{5px}

%% file: sections/Tables/Features.tex
\begin{table*}[t]
    \caption{Features used by the \classifier}
\label{table:Features}
\vspace{-5pt}
\newcommand{\failureRateWidth}{2.5in}
\newcommand{\failureRateHeight}{4em}
\scriptsize
\centering
    \begin{tabular}{l|c|l}
    \toprule     
     \textbf{Feature Name}&\textbf{Type}&\textbf{Description}\\
        \midrule
        Exception Type & Str & The name of the exception e.g. UnknownHostException \\
        Test name in Stacktrace & Boolean & \textit{True} if one of Stacktrace lines starts with the test name else \textit{False} \\
        Test Class name in Stacktrace & Boolean & \textit{True} if one of Stacktrace lines contains the test class name else \textit{False} \\
        Other Tests in Stacktrace & Boolean & \textit{True} if one of Stacktrace lines starts with other tests names else \textit{False} \\
        JUnit in Stacktrace & Boolean & \textit{True} if one of Stacktrace lines starts with any Junit Lines else \textit{False} \\
        CUT in Stacktrace & Boolean & \textit{True} if one of Stacktrace lines contains any lines from Code Under Test else \textit{False} \\
\bottomrule 
\end{tabular}
\vspace{-10pt}
\end{table*}

%% file: sections/FinalStyle/evaluation.tex
\section{Evaluation Methodology}
\label{sec:evaluation}
The core contribution in this work is a rigorous empirical evaluation of the three flaky failure detection approaches described in the prior section, using the following methodology:  

 \subsection{Datasets}

In order to effectively evaluate failure de-duplication for flaky failures, we need a dataset that contains a large number of both flaky and true failures for the same test.
The``FlakeFlagger'' dataset was built by executing the test suites of 26 open-source Java projects 10,000 times and recording their outputs, yielding a large dataset of flaky failures~\cite{alshammari2021flakeflagger}.
We choose the FlakeFlagger dataset, as it contains the complete failure logs for each flaky failure, as opposed to other flaky test datasets like DeFlaker's~\cite{bell2018deflaker} or iDFlakies~\cite{lam2019idflakies}.

Whereas a dataset of flaky failures can be mined by repeatedly running the same versions of the same tests, a dataset of true failures can only be mined from buggy code.
While datasets of true failures \emph{do} exist \cite{just2014defects4j,saha2018bugs,tomassi2019bugswarm,bears}, these datasets are typically intentionally constructed from tests that are \emph{not} flaky (to make studying the defects easier).
However, we are not aware of any accessible datasets that provide both flaky and true failures logs for the same set of tests.
Even if one were to mine failures of flaky tests, there would still be a tremendous dataset imbalance problem: there tend to be far more tests that only fail due to flakiness as opposed to those that might also reveal faults~\cite{haben2023importance}.

We propose a novel methodology for constructing a dataset for this experiment, based on mutation testing.
Mutation testing runs a program's test suite on generated mutants (variants of the program under test), and evaluates how many of those mutants are detected by a failing test.
Mutants have been shown to be an effective substitute for real faults in software testing~\cite{just2014mutants}.
Hence, for each of the flaky tests in our dataset, we use mutation testing to build a large dataset of failure logs for true failures.
To avoid contaminating the true failure dataset with flaky failures (caused by tests failing due to flakiness on the mutated code), we apply Shi et al.'s approach for filtering flaky mutants~\cite{shi2019mitigating}.

Hence, the dataset for our experiment consists of all of the flaky failures extracted from the FlakeFlagger dataset~\cite{alshammari2021flakeflagger}, supplemented by true failures generated by executing Shi et al.'s version of the popular PIT mutation testing tool~\cite{shi2019mitigating,coles2016pit}.
This modified version of PIT is configured such that each test-mutant failure is confirmed by re-running the test on that mutant, 20 times.
Each failure that is deterministically reproduced is included in our dataset of failures.
This confirmation step is necessary to filter out any flaky failures from the mutation dataset, and is used only for confirming that the failure is deterministic (we do not include each failure 20 times from each of the confirmation runs).
Then from the collected failure logs of each killed mutant, we collect the \failures. We extend the XML file per test to include a list of killed mutants, each of them contains the \failure. 

In practice, flaky failures tend to be far more common than true failures.
Given that the \failure includes the name of each test, the performance of any failure classifier could be misrepresented by a dataset that contained a large proportion of tests that \emph{only} failed due to flakiness.
For example, in a 9-month period observing Google's Chromium CI, Haben et al. observed that 1,446 tests failed with only true (``fault-revealing'') failures, 22,477 failed with only flaky failures, and 897 failed showing both failures.
A predictor based on the historical flaky failure rate of a test would easily have quite high recall at predicting flakiness (e.g. having at most $897/22,477=4\%$ true failures incorrectly labeled as flaky). 
Our goal is to evaluate the performance of approaches that rely primarily on the \failure, and \emph{not} just the historical flake rate of a test.

Hence, we include in our evaluation \emph{only} tests with at least one flaky and non-flaky failure, and report the number of true and flaky failures in our dataset for each project.
We were not able to successfully apply the PIT mutation testing tool to all of the projects despite significant efforts (one author expended at least 2 hours per-project to attempt to get it to work) --- and hence, we were unable to gather a resource of failures for all projects.
As a result, it is important to note that we do \emph{not} include all projects or tests from the FlakeFlagger dataset.
Whereas the FlakeFlagger dataset includes 811 flaky tests from 24 projects, we analyze only those tests for which we could collect a dataset of true failures: 498 flaky tests from 22 projects.
For example: on the project ``spring-boot,'' the dataset contained 163 flaky tests, but we were only able to successfully run PIT on 12 of the tests.
The errors were primarily related to interactions between bytecode instrumentation, classloading, and custom JUnit runners.

To evaluate our classifier, we use cross validation to split the whole dataset (flaky and true failures) to a training set and testing set.
As the dataset is imbalanced (with different proportions of failures that are flaky vs true failures), we  apply SMOTE~\cite{SMOTE} when training and utilize stratified cross-validation~\cite{crossValidation} to ensure that each testing-fold part has at least one flaky failure.

\subsection{Research Questions}

Using this dataset of 498 tests with both flaky and true failures, we design an experiment to answer the following research questions:

\begin{description}
    \item \textbf{RQ1: How often are flaky failures repetitive?} 
    We examine how frequently a flaky failure matches \emph{at least} one other flaky failure --- of the same test or of other tests within the same project. By doing this, we show the repetition of flaky failures and the efficacy of the failure de-duplication approach. 
    
    \item \textbf{RQ2: With prior flaky and true failures, is it feasible to use failure de-duplication to determine if a failure is flaky or true one?}
    The main objective is to evaluate the effectiveness of using \syntax as an approach to find the differences between flaky and true failures. This helps practitioners and researchers if they can rely on the approach in detecting flaky failures. Since projects differ in their domain, root causes of flakiness, and the total number of flaky tests, we evaluate the approach on a project-by-project basis.

    \item\textbf{RQ3: To what extent is the utilization of machine learning helpful in finding the differences between flaky and true failures?} We aim to demonstrate the efficacy of employing machine learning classifiers in predicting whether a failure is flaky or not based on specific features extracted from failure logs. We examine whether a classifier can leverage failures from other tests within the same project to enhance the learning process of the model to better predict failures. 
    
\end{description}

\section{Results}
\input{sections/FinalStyle/RQs/RQ1}

\input{sections/FinalStyle/RQs/RQ2}

\input{sections/FinalStyle/RQs/RQ3}

%% file: sections/FinalStyle/RQs/RQ1.tex
\subsection{\textbf{RQ1: How often are flaky failures repetitive?}}
\label{sec:rq1}

\input{sections/Tables/repetitiveeFailures}

We apply the \syntax approach to de-duplicate flaky failures, and summarize the results in Table~\ref{table:repetitive}.
We show results matching the flaky failures within the same test (shown in column \emph{Per Test}), and matching the flaky failures across all failures from all tests in the same project (shown in the column \emph{Across Tests}). 
By considering \alluxio as an example from Table~\ref{table:repetitive}: in the first case, there are 114 flaky tests and those tests cumulatively have 16,858 flaky failures in total (16,847 of them are repetitive and 11 are not). 
The 16,847 failures that were an exact match for at least one other failure represent just 310 unique failures.
In the second case, the number of failures that are \emph{not} repetitive dropped to 5 (16,853 repetitive failures).  
When comparing a new failure to flaky failures from different tests, the \syntax might produce mis-match results due to lines in the stacktrace pointing to the test. To mitigate this, we exclude such lines during this type of comparison, ensuring a more accurate match result.

While we found that each flaky test in \alluxio could have different flaky failures, on average, each flaky test only had just over two different failures, each of which recurred many times.
In most of the projects we studied in Table~\ref{table:repetitive}, there are a reasonable amount of repetitive flaky failures (by both considering the ratio of the number of flaky failures in column (\textbf{Uniq}) to the total number of flaky failures or even to the set of flaky failures) as some projects (8 out of 22) have all flaky failures repetitive.
Hence, we conclude that, overall, flaky failures are extremely repetitive.
While it is inappropriate to assume that each flaky test can only fail with a single set of symptoms, the number of unique failures is dwarfed by the frequency with which those failures recur.

We also carefully examine when flaky failures are not repetitive, and occur only once in the dataset.
Across all the studied projects, there are only 123 out of 80,530 flaky failures (also out of 763 sets of flaky failures) that have never matched other flaky failures within the same project.
Out of 123 that failed once, we found 90 of them are actually lack of the history of flaky failures (from tests that only failed once). Out of 22 projects, there are only two projects where the number of repetitive flaky failure is just equal or less than the number of non-repetitive flaky failures (\elastic and \spring), and all these failures (except one in \elastic) are from tests that only fail once.

\input{sections/Tables/non-unique-failures-result}

While it is common for frequently failing flaky tests to exhibit repetitive flaky failures, this trend is not consistent across all projects. For example, within the project \hbase, there are 5 flaky tests that failed more than 1,000 times have at least one non-repetitive flaky failure. 

We investigated whether specific exception types were associated with these non-repetitive flaky failures. From the dataset we analyzed, among the top 10 most frequently occurring exceptions, two exceptions appeared more frequently in non-repetitive failures than repetitive flaky failures. Specifically, the \emph{RuntimeException} was observed 13 times out of its 22 failure cases, while the \emph{SocketException} was also observed 19 times (out of 31 failures). We found that every failures with the \emph{SocketException} was linked within the \okhttp project.

We observed that certain test suite runs, especially those with a higher number of failed tests, tend to exhibit repetitive flaky failures across most or all the failed tests. For instance, within the \ambari, 47 out of 51 flaky tests consistently failed together and displayed the same \failures each time they failed, and none of their stacktrace lines contain the test names.

\textbf{Summary}. Flaky failures are often repetitive. This can serve as an indicator for developers: previous flaky failures can be a reference to check if a newly encountered failure is familiar. However, there are \emph{few} cases where a failure is not similar with any previously observed flaky failures. In such situations, a deeper investigation is needed to detect its flakiness. A valuable step in this investigative process involves comparing the failure with flaky failures from other tests, especially when the failure's stacktrace lines do not reference the test itself.

%% file: sections/Tables/repetitiveeFailures.tex
\begin{table}[t]
  \setlength{\tabcolsep}{2pt}

\caption{Repetitive Flaky Failures within and across tests per project. \\
\textnormal{Failures column shows the number of flaky failures and the different failures (Set). The columns \emph{Repet} and \emph{Uniq} refer to flaky failures that are and are not repetitive, respectively.  Per Test refers to matching the failures within the same test. Across Tests refers matching all flaky failures from all tests.}
}
\label{table:repetitive}
\vspace{-4pt}
\footnotesize
\begin{tabular}{l|r|rr|rr|rr}

\toprule
      & &\multicolumn{2}{c|}{\textbf{Failures}} & \multicolumn{2}{c|}{\textbf{Per Test}} & \multicolumn{2}{c}{\textbf{Across Tests}}\\

\textbf{Projects} & \textbf{Tests} & \textbf{Flaky}  & \textbf{Set}  & \textbf{Uniq} & \textbf{Repet}  & \textbf{Uniq} & \textbf{Repet} \\
\midrule


Alluxio-alluxio&114&16,858&310&11&16,847&5&16,853\\
\cellcolor{gray!6}{square-okhttp}&\cellcolor{gray!6}{99}&\cellcolor{gray!6}{26,486}&\cellcolor{gray!6}{120}&\cellcolor{gray!6}{40}&\cellcolor{gray!6}{26,446}&\cellcolor{gray!6}{17}&\cellcolor{gray!6}{26,469}\\
apache-ambari&51&4,063&54&0&4,063&0&4,063\\
\cellcolor{gray!6}{hector-client-hector}&\cellcolor{gray!6}{33}&\cellcolor{gray!6}{6,529}&\cellcolor{gray!6}{33}&\cellcolor{gray!6}{0}&\cellcolor{gray!6}{6,529}&\cellcolor{gray!6}{0}&\cellcolor{gray!6}{6,529}\\
activiti-activiti&30&1,363&31&13&1,350&6&1,357\\
\cellcolor{gray!6}{tootallnate-java-websocket}&\cellcolor{gray!6}{23}&\cellcolor{gray!6}{2,143}&\cellcolor{gray!6}{45}&\cellcolor{gray!6}{2}&\cellcolor{gray!6}{2,141}&\cellcolor{gray!6}{0}&\cellcolor{gray!6}{2,143}\\
apache-httpcore&22&354&22&9&345&2&352\\
\cellcolor{gray!6}{qos-ch-logback}&\cellcolor{gray!6}{20}&\cellcolor{gray!6}{438}&\cellcolor{gray!6}{21}&\cellcolor{gray!6}{8}&\cellcolor{gray!6}{430}&\cellcolor{gray!6}{4}&\cellcolor{gray!6}{434}\\
apache-hbase&20&2,519&26&3&2,516&2&2,517\\
\cellcolor{gray!6}{kevinsawicki.http-request}&\cellcolor{gray!6}{18}&\cellcolor{gray!6}{3,501}&\cellcolor{gray!6}{18}&\cellcolor{gray!6}{3}&\cellcolor{gray!6}{3,498}&\cellcolor{gray!6}{0}&\cellcolor{gray!6}{3,501}\\
wildfly-wildfly&18&50&18&12&38&4&46\\
\cellcolor{gray!6}{wro4j-wro4j}&\cellcolor{gray!6}{14}&\cellcolor{gray!6}{10,833}&\cellcolor{gray!6}{21}&\cellcolor{gray!6}{3}&\cellcolor{gray!6}{10,830}&\cellcolor{gray!6}{2}&\cellcolor{gray!6}{10,831}\\
spring-projects-spring-boot&12&14&13&12&2&5&9\\
\cellcolor{gray!6}{undertow-io-undertow}&\cellcolor{gray!6}{7}&\cellcolor{gray!6}{92}&\cellcolor{gray!6}{12}&\cellcolor{gray!6}{3}&\cellcolor{gray!6}{89}&\cellcolor{gray!6}{1}&\cellcolor{gray!6}{91}\\
orbit-orbit&7&2,943&7&0&2,943&0&2,943\\
\cellcolor{gray!6}{elasticjob-elastic-job-lite}&\cellcolor{gray!6}{3}&\cellcolor{gray!6}{7}&\cellcolor{gray!6}{4}&\cellcolor{gray!6}{3}&\cellcolor{gray!6}{4}&\cellcolor{gray!6}{0}&\cellcolor{gray!6}{7}\\
doanduyhai-Achilles&2&121&3&1&120&1&120\\
\cellcolor{gray!6}{joel-costigliola-assertj-core}&\cellcolor{gray!6}{1}&\cellcolor{gray!6}{974}&\cellcolor{gray!6}{1}&\cellcolor{gray!6}{0}&\cellcolor{gray!6}{974}&\cellcolor{gray!6}{0}&\cellcolor{gray!6}{974}\\
ninjaframework-ninja&1&476&1&0&476&0&476\\
\cellcolor{gray!6}{apache-commons-exec}&\cellcolor{gray!6}{1}&\cellcolor{gray!6}{33}&\cellcolor{gray!6}{1}&\cellcolor{gray!6}{0}&\cellcolor{gray!6}{33}&\cellcolor{gray!6}{0}&\cellcolor{gray!6}{33}\\
jknack-handlebars.java&1&411&1&0&411&0&411\\
\cellcolor{gray!6}{zxing-zxing}&\cellcolor{gray!6}{1}&\cellcolor{gray!6}{322}&\cellcolor{gray!6}{1}&\cellcolor{gray!6}{0}&\cellcolor{gray!6}{322}&\cellcolor{gray!6}{0}&\cellcolor{gray!6}{322}\\
\midrule
Total&498&80,530&763&123&80,407&49&80,481\\

\bottomrule
\end{tabular}
\vspace{-10pt}
\end{table}

%% file: sections/Tables/non-unique-failures-result.tex
\begin{table*}[t]
\caption{Text-Based matching to label flaky and true (non-flaky) failures.
\textnormal{The \textit{Total Tests and Failures} column provides the total flaky tests, the number of true (non-flaky) failures across these tests, and the count of flaky failures.  
The \textit{Set of Failures} column displays the different failures within both flaky and true failures.
The \textit{Confusion Matrix and Evaluation By Failures} columns present the matching results between flaky and true failures. The \# of Tests in TP and FN shows how many different tests have at least one failure in each category. The cumulative number of tests in \emph{TP} and \emph{FN} might exceed the total given in \emph{Test} because a test might have multiple flaky failures in different categories.
}
}

\vspace{-5pt}
\setlength{\tabcolsep}{5.0pt}
\newcommand{\failureRateWidth}{2.5in}
\newcommand{\failureRateHeight}{4em}
\scriptsize
\centering

    \begin{tabular}{l|rrr|rr|rrrrrrr|rr}
    \toprule
      
      & \multicolumn{3}{c|}{\textbf{Total Tests and Failures}} & \multicolumn{2}{c|}{\textbf{Set of Failures}} & \multicolumn{7}{c}{\textbf{Confusion Matrix and Evaluation By Failures}} &  \multicolumn{2}{c}{\textbf{\# of Tests in}} \\
    \midrule
     \textbf{Project}&\textbf{Tests}&\textbf{Non-Flaky}&\textbf{Flaky} &\textbf{True}&\textbf{Flaky} &\textbf{TP}&\textbf{FN}&\textbf{FP}&\textbf{TN}&\textbf{P} &\textbf{R} &\textbf{SP} & \textbf{TP}&\textbf{FN}\\
\midrule

Alluxio-alluxio&114&32,795&16,858&6,232&310&9,173&7,685&1,933&30,862&82\%&54\%&94\%&114&109\\
\cellcolor{gray!6}{square-okhttp}&\cellcolor{gray!6}{99}&\cellcolor{gray!6}{33,949}&\cellcolor{gray!6}{26,486}&\cellcolor{gray!6}{18,546}&\cellcolor{gray!6}{120}&\cellcolor{gray!6}{16,517}&\cellcolor{gray!6}{9,969}&\cellcolor{gray!6}{107}&\cellcolor{gray!6}{33,842}&\cellcolor{gray!6}{99\%}&\cellcolor{gray!6}{62\%}&\cellcolor{gray!6}{99\%}&\cellcolor{gray!6}{58}&\cellcolor{gray!6}{52}\\
apache-ambari&51&11,045&4,063&4,562&54&4,003&60&5&11,040&99\%&98\%&99\%&50&2\\
\cellcolor{gray!6}{hector-client-hector}&\cellcolor{gray!6}{33}&\cellcolor{gray!6}{3,603}&\cellcolor{gray!6}{6,529}&\cellcolor{gray!6}{1,769}&\cellcolor{gray!6}{33}&\cellcolor{gray!6}{1,382}&\cellcolor{gray!6}{5,147}&\cellcolor{gray!6}{12}&\cellcolor{gray!6}{3,591}&\cellcolor{gray!6}{99\%}&\cellcolor{gray!6}{21\%}&\cellcolor{gray!6}{99\%}&\cellcolor{gray!6}{32}&\cellcolor{gray!6}{1}\\
activiti-activiti&30&44,937&1,363&15,863&31&932&431&2,272&42,665&29\%&68\%&94\%&1&29\\
\cellcolor{gray!6}{tootallnate-java-websocket}&\cellcolor{gray!6}{23}&\cellcolor{gray!6}{1,116}&\cellcolor{gray!6}{2,143}&\cellcolor{gray!6}{418}&\cellcolor{gray!6}{45}&\cellcolor{gray!6}{596}&\cellcolor{gray!6}{1,547}&\cellcolor{gray!6}{531}&\cellcolor{gray!6}{585}&\cellcolor{gray!6}{52\%}&\cellcolor{gray!6}{27\%}&\cellcolor{gray!6}{52\%}&\cellcolor{gray!6}{20}&\cellcolor{gray!6}{23}\\
apache-httpcore&22&8,021&354&667&22&0&354&2,096&5,925&0\%&0\%&73\%&0&22\\
\cellcolor{gray!6}{apache-hbase}&\cellcolor{gray!6}{20}&\cellcolor{gray!6}{585}&\cellcolor{gray!6}{2,519}&\cellcolor{gray!6}{185}&\cellcolor{gray!6}{26}&\cellcolor{gray!6}{1,209}&\cellcolor{gray!6}{1,310}&\cellcolor{gray!6}{162}&\cellcolor{gray!6}{423}&\cellcolor{gray!6}{88\%}&\cellcolor{gray!6}{47\%}&\cellcolor{gray!6}{72\%}&\cellcolor{gray!6}{17}&\cellcolor{gray!6}{5}\\
qos-ch-logback&20&2,614&438&895&21&56&382&368&2,246&13\%&12\%&85\%&3&17\\
\cellcolor{gray!6}{kevinsawicki.http-request}&\cellcolor{gray!6}{18}&\cellcolor{gray!6}{387}&\cellcolor{gray!6}{3,501}&\cellcolor{gray!6}{229}&\cellcolor{gray!6}{18}&\cellcolor{gray!6}{981}&\cellcolor{gray!6}{2,520}&\cellcolor{gray!6}{40}&\cellcolor{gray!6}{347}&\cellcolor{gray!6}{96\%}&\cellcolor{gray!6}{28\%}&\cellcolor{gray!6}{89\%}&\cellcolor{gray!6}{4}&\cellcolor{gray!6}{14}\\
wildfly-wildfly&18&4,364&50&1,497&18&38&12&0&4,364&100\%&76\%&100\%&6&12\\
\cellcolor{gray!6}{wro4j-wro4j}&\cellcolor{gray!6}{14}&\cellcolor{gray!6}{540}&\cellcolor{gray!6}{10,833}&\cellcolor{gray!6}{90}&\cellcolor{gray!6}{21}&\cellcolor{gray!6}{800}&\cellcolor{gray!6}{10,033}&\cellcolor{gray!6}{29}&\cellcolor{gray!6}{511}&\cellcolor{gray!6}{96\%}&\cellcolor{gray!6}{7\%}&\cellcolor{gray!6}{94\%}&\cellcolor{gray!6}{9}&\cellcolor{gray!6}{11}\\
spring-projects-spring-boot&12&2,150&14&244&13&2&12&0&2,150&100\%&14\%&100\%&1&12\\
\cellcolor{gray!6}{undertow-io-undertow}&\cellcolor{gray!6}{7}&\cellcolor{gray!6}{2,306}&\cellcolor{gray!6}{92}&\cellcolor{gray!6}{236}&\cellcolor{gray!6}{12}&\cellcolor{gray!6}{8}&\cellcolor{gray!6}{84}&\cellcolor{gray!6}{943}&\cellcolor{gray!6}{1,363}&\cellcolor{gray!6}{0\%}&\cellcolor{gray!6}{8\%}&\cellcolor{gray!6}{59\%}&\cellcolor{gray!6}{2}&\cellcolor{gray!6}{6}\\
orbit-orbit&7&812&2,943&302&7&87&2,856&57&755&60\%&2\%&92\%&2&5\\
\cellcolor{gray!6}{elasticjob-elastic-job-lite}&\cellcolor{gray!6}{3}&\cellcolor{gray!6}{111}&\cellcolor{gray!6}{7}&\cellcolor{gray!6}{68}&\cellcolor{gray!6}{4}&\cellcolor{gray!6}{4}&\cellcolor{gray!6}{3}&\cellcolor{gray!6}{0}&\cellcolor{gray!6}{111}&\cellcolor{gray!6}{100\%}&\cellcolor{gray!6}{57\%}&\cellcolor{gray!6}{100\%}&\cellcolor{gray!6}{1}&\cellcolor{gray!6}{3}\\
doanduyhai-Achilles&2&154&121&86&3&120&1&6&148&95\%&99\%&96\%&1&1\\
\cellcolor{gray!6}{jknack-handlebars.java}&\cellcolor{gray!6}{1}&\cellcolor{gray!6}{147}&\cellcolor{gray!6}{411}&\cellcolor{gray!6}{61}&\cellcolor{gray!6}{1}&\cellcolor{gray!6}{0}&\cellcolor{gray!6}{411}&\cellcolor{gray!6}{16}&\cellcolor{gray!6}{131}&\cellcolor{gray!6}{0\%}&\cellcolor{gray!6}{0\%}&\cellcolor{gray!6}{89\%}&\cellcolor{gray!6}{0}&\cellcolor{gray!6}{1}\\
zxing-zxing&1&76&322&37&1&322&0&0&76&100\%&100\%&100\%&1&0\\
\cellcolor{gray!6}{joel-costigliola-assertj-core}&\cellcolor{gray!6}{1}&\cellcolor{gray!6}{18}&\cellcolor{gray!6}{974}&\cellcolor{gray!6}{10}&\cellcolor{gray!6}{1}&\cellcolor{gray!6}{974}&\cellcolor{gray!6}{0}&\cellcolor{gray!6}{0}&\cellcolor{gray!6}{18}&\cellcolor{gray!6}{100\%}&\cellcolor{gray!6}{100\%}&\cellcolor{gray!6}{100\%}&\cellcolor{gray!6}{1}&\cellcolor{gray!6}{0}\\
apache-commons-exec&1&59&33&13&1&0&33&2&57&0\%&0\%&96\%&0&1\\
\cellcolor{gray!6}{ninjaframework-ninja}&\cellcolor{gray!6}{1}&\cellcolor{gray!6}{120}&\cellcolor{gray!6}{476}&\cellcolor{gray!6}{4}&\cellcolor{gray!6}{1}&\cellcolor{gray!6}{0}&\cellcolor{gray!6}{476}&\cellcolor{gray!6}{8}&\cellcolor{gray!6}{112}&\cellcolor{gray!6}{0\%}&\cellcolor{gray!6}{0\%}&\cellcolor{gray!6}{93\%}&\cellcolor{gray!6}{0}&\cellcolor{gray!6}{1}\\
\midrule
22 Projects Total&498&149,909&80,530&52,014&763&37,204&43,326&8,587&141,322&&&&323&327\\

\bottomrule
\end{tabular}

\label{nonunique}
\vspace{-10pt}
\end{table*}

%% file: sections/FinalStyle/RQs/RQ2.tex
\subsection{\textbf{RQ2: With prior flaky and true failures, is it feasible to use the failure de-duplication to tell if a failure is flaky or true one?}}

We investigate if the \syntax can be used to determine if a failure is flaky or not based on the failure de-duplication. As we consider both flaky and true failures, we use a confusion matrix as follows:

\begin{description}
    \item \textbf{TP}: Flaky failures that match at least one flaky failure and do not match any of the true failures. 
    \item \textbf{FN}: Flaky failures that match at least one true failure \textit{or} does not match with any of the flaky failures. 
    \item \textbf{FP}: True failures that match at least one flaky failure. 
    \item \textbf{TN}: True failures that do not match with any of flaky failure.
    
\end{description}

This evaluation methodology follows our running use-case, where newly observed test failures are either labeled as flaky (and ignored), or triaged to developers for further debugging and analysis.
We then evaluate the result of matching using the \emph{Precision} (\textbf{P}), \emph{Recall} (\textbf{R}),and \emph{Specificity} (\textbf{SP}) as follows:

\begin{equation}
\label{precision}
\text{Precision (\textbf{P})} = \frac{\text{TP}}{\text{TP + FP}}
\end{equation}

\begin{equation}
\label{recall}
\text{Recall (\textbf{R})} = \frac{\text{TP}}{\text{TP + FN}}
\end{equation}

\begin{equation}
\label{specificity}
\text{Specificity (\textbf{SP})} = \frac{\text{TN}}{\text{TN + FP}}
\end{equation}

We chose these metrics to reflect the use-case of developers encountering a failure and comparing it with historical flaky and true failures.
Given a model where developers ignore test failures that are labeled as flaky, a safer approach would have a higher precision, as precision reports the frequency with which an approach falsely determines a test to be flaky.
Since we consider scenarios where developers may be most interested in minimizing false positives, we also report specificity, which evaluates the percentage of true failures correctly labeled. 
Lower recall scores indicate that an approach inadvertently labels more flaky failures as true failures --- indicating that a developer might spend more time debugging them.

\input{sections/Tables/exceptionsTable}

Table~\ref{nonunique} shows the confusion matrix of using our approach as described in Section~\ref{sec:evaluation}.
The performance of the approach varies across projects. For example, there are projects with at least 95\% precision (10 out of 22) while some projects with 0\% (5 out 10 projects).

We examine the results per-project to gain further insights into the performance of the approach.
We find that in projects where the \syntax approach struggles to differentiate between flaky and true failures, failures are typically presented as \emph{assertion} exceptions. 
For example: \emph{all} of the false negative (\emph{FN}) flaky failures in \websocket, \emph{all} of the \emph{FN} flaky failures in \orbit, and 98\% of the \emph{FN} in \okhttp are assertion exceptions.
In the case of \alluxio, we found that 90\% of the \emph{FN} failures were \emph{NullPointerException}s.
Even with the availability of stacktraces in these failures, these exceptions remain challenging to be used in finding the differences between flaky and true failures. 
On the other side, the projects which have reasonable precision and recall scores (or at least precision scores) like the case in \Achilles, there are a verity of different exceptions like \emph{UnknownHostException}, and less likely to have general exceptions such as \emph{assertion} and \emph{NullPointerException}.

We also examined the relationship between the performance of the approach and factors such as the proportion of true failures and the number of times that a test flakes.
Examining the table, we see two projects with a comparable ratio between flaky and non-flaky projects: \httpcore and \wildfly.
However, we also note that the performance of the approach (particularly in terms of true positives) varies significantly between these two projects.
Overall, we do not note any significant correlation between the ratio of flaky to true failures and the performance of the approach.

Examining projects with more than one flaky test, we found the project \ambari has the best performance in \emph{precision}, \emph{recall}, and \emph{specificity}.
In this project, we found that the majority of the flaky failures failed with the exception \emph{ProvisionException}.
As discussed in RQ1 (Section~\ref{sec:rq1}), the majority of flaky tests in this project failed together, in the same test suite execution.
This case is somewhat different from the other projects, in which flaky failures occur in different test runs --- it is indeed quite likely that all of the flaky failures have the same root cause.

We also show the number of tests with at least one true positive or false negative failure (the last two columns of Table~\ref{nonunique}).
For example, in the case of \alluxio, we see that of 114 tests in total, all 114 had at least one failure that was correctly classified as a true positive.
However, 109 of those 114 tests \emph{also} had at least one failure falsely classified as a false negative (falsely labeled as ``not flaky'').
These data indicate that na\"{i}ve approaches that rely on test name to determine whether or not a failure is flaky or not are unlikely to be effective on this dataset.

To gain insight into the value of matching stack traces (in addition to exceptions), we also examined the performance of matching failures \emph{only} using exception type.
Table~\ref{table:exceptions} presents the top ten most frequently occurring exceptions observed in the analyzed flaky failures.
In some cases, the exception by itself cannot determine the differences of matching compared the case when we consider the stacktraces such as the case of \emph{NullPointerException}.
However, we see that \emph{UnknownHostException} occurred almost exclusively in the context of flaky failures.
However, this certainly does \emph{not} support the conclusion that all instances of this exception in a failure indicate a flaky failure: in three other projects in which this exception occurred, it occurred only in true failures. 
This suggests that some exceptions could be linked to flakiness within a project, but it is likely not possible to generalize this correlation across projects.

We also discovered some failures where exceptions match both flaky and true failures, as indicated in Table~\ref{table:exceptions}. In the context of our experiment, the most frequently occurring exception is the \emph{AssertionError} which roughly 20\% of these failures appear in \textbf{TP}. However, when considering only the exception type and excluding stacktrace lines, the proportion drops to less than 1\%.
The reason behind this observation is the generality of the \emph{AssertionError} exception. For example, a test may have multiple assertion statements, and if they fail for different reasons, they match the exception but differ in the stacktrace. Therefore, it becomes challenging to attribute this type of exception to a specific type of failure.

\textbf{Summary}. We found that using the de-duplication approach to find flaky and true failures effective in some projects especially when their failures logs more informative than just assertion failures. For most of the projects, relying on the stacktraces in addition to the exception type is helpful as most failure exceptions could be seen in both flaky and true failures. The approach performs best when failure messages are specific.

%% file: sections/Tables/exceptionsTable.tex
\begin{table*}[t]
\caption{Top 10 Most Occurrence Exception in Flaky and True Failures \\ 
\textnormal{ The \textit{Exception Occurrence} column details the frequency of a specific exception, indicating in how many projects, tests, and failures this exception has been observed. The \textit{Match Result (with Stacktraces)} column displays the match distributions, considering stacktraces and the related test count while the, \textit{Match Result (without Stacktraces)} column indicates match results based on exception types, excluding stacktraces.}}

\vspace{-5pt}
\setlength{\tabcolsep}{2.0pt}
\newcommand{\failureRateWidth}{2.5in}
\newcommand{\failureRateHeight}{4em}
\scriptsize
\centering

    \begin{tabular}{l|rrr|rr|rrrr|rrrr|rrrr|rrrr}
    \toprule
      & \multicolumn{5}{c|}{\textbf{Exception Occurrence}} & \multicolumn{8}{c|}{\textbf{Match Result (with Stacktraces)}} & \multicolumn{8}{c}{\textbf{Match Result (without Stacktraces)}} \\ 

      & \multicolumn{5}{c|}{\textbf{}} & \multicolumn{4}{c|}{\textbf{By Failures}} & \multicolumn{4}{c|}{\textbf{By Tests}} & \multicolumn{4}{c}{\textbf{By Failures}} & \multicolumn{4}{c}{\textbf{By Tests}} \\ 
     
     \textbf{Exception Name}&\textbf{Projects}&\textbf{Tests}&\textbf{Failures}&\textbf{True}&\textbf{Flaky}&\textbf{TP}& \textbf{FN}&\textbf{FP}& \textbf{TN}&\textbf{TP}& \textbf{FN}&\textbf{FP}& \textbf{TN}&\textbf{TP}& \textbf{FN}&\textbf{FP}& \textbf{TN}&\textbf{TP}& \textbf{FN}&\textbf{FP}& \textbf{TN}\\
        \midrule

NullPointerException&21&475&50,427&42,264&8,163&1,168&6,995&711&41,553&40&106&105&475&0&8,163&8,444&33,820&0&120&120&355\\
\cellcolor{gray!6}{AssertionError}&\cellcolor{gray!6}{21}&\cellcolor{gray!6}{401}&\cellcolor{gray!6}{48,751}&\cellcolor{gray!6}{19,727}&\cellcolor{gray!6}{29,024}&\cellcolor{gray!6}{5,977}&\cellcolor{gray!6}{23,047}&\cellcolor{gray!6}{4,376}&\cellcolor{gray!6}{15,351}&\cellcolor{gray!6}{63}&\cellcolor{gray!6}{120}&\cellcolor{gray!6}{94}&\cellcolor{gray!6}{367}&\cellcolor{gray!6}{253}&\cellcolor{gray!6}{28,771}&\cellcolor{gray!6}{13,191}&\cellcolor{gray!6}{6,536}&\cellcolor{gray!6}{4}&\cellcolor{gray!6}{174}&\cellcolor{gray!6}{171}&\cellcolor{gray!6}{223}\\
IOException&7&219&16,959&15,800&1,159&642&517&515&15,285&10&16&13&206&642&517&3,718&12,082&10&16&13&193\\
\cellcolor{gray!6}{UnknownHostException}&\cellcolor{gray!6}{9}&\cellcolor{gray!6}{225}&\cellcolor{gray!6}{9,369}&\cellcolor{gray!6}{315}&\cellcolor{gray!6}{9,054}&\cellcolor{gray!6}{9,052}&\cellcolor{gray!6}{2}&\cellcolor{gray!6}{0}&\cellcolor{gray!6}{315}&\cellcolor{gray!6}{125}&\cellcolor{gray!6}{2}&\cellcolor{gray!6}{0}&\cellcolor{gray!6}{98}&\cellcolor{gray!6}{9,052}&\cellcolor{gray!6}{2}&\cellcolor{gray!6}{0}&\cellcolor{gray!6}{315}&\cellcolor{gray!6}{125}&\cellcolor{gray!6}{2}&\cellcolor{gray!6}{0}&\cellcolor{gray!6}{98}\\
ActivitiException&1&30&9,277&9,205&72&0&72&342&8,863&0&9&7&29&0&72&2,809&6,396&0&9&8&21\\
\cellcolor{gray!6}{IllegalArgumentException}&\cellcolor{gray!6}{17}&\cellcolor{gray!6}{393}&\cellcolor{gray!6}{8,666}&\cellcolor{gray!6}{8,663}&\cellcolor{gray!6}{3}&\cellcolor{gray!6}{0}&\cellcolor{gray!6}{3}&\cellcolor{gray!6}{189}&\cellcolor{gray!6}{8,474}&\cellcolor{gray!6}{0}&\cellcolor{gray!6}{3}&\cellcolor{gray!6}{3}&\cellcolor{gray!6}{393}&\cellcolor{gray!6}{0}&\cellcolor{gray!6}{3}&\cellcolor{gray!6}{203}&\cellcolor{gray!6}{8,460}&\cellcolor{gray!6}{0}&\cellcolor{gray!6}{3}&\cellcolor{gray!6}{3}&\cellcolor{gray!6}{390}\\
AssertionFailedError&7&91&8,644&6,881&1,763&66&1,697&1,598&5,283&1&20&20&88&66&1,697&4,074&2,807&1&20&20&70\\
\cellcolor{gray!6}{NoSuchMethodError}&\cellcolor{gray!6}{1}&\cellcolor{gray!6}{1}&\cellcolor{gray!6}{8,539}&\cellcolor{gray!6}{0}&\cellcolor{gray!6}{8,539}&\cellcolor{gray!6}{8,539}&\cellcolor{gray!6}{0}&\cellcolor{gray!6}{0}&\cellcolor{gray!6}{0}&\cellcolor{gray!6}{1}&\cellcolor{gray!6}{0}&\cellcolor{gray!6}{0}&\cellcolor{gray!6}{0}&\cellcolor{gray!6}{8,539}&\cellcolor{gray!6}{0}&\cellcolor{gray!6}{0}&\cellcolor{gray!6}{0}&\cellcolor{gray!6}{1}&\cellcolor{gray!6}{0}&\cellcolor{gray!6}{0}&\cellcolor{gray!6}{0}\\
PersistenceException&2&30&8,399&8,398&1&0&1&156&8,242&0&1&1&30&0&1&389&8,009&0&1&1&29\\
\cellcolor{gray!6}{WroRuntimeException}&\cellcolor{gray!6}{1}&\cellcolor{gray!6}{7}&\cellcolor{gray!6}{6,520}&\cellcolor{gray!6}{33}&\cellcolor{gray!6}{6,487}&\cellcolor{gray!6}{0}&\cellcolor{gray!6}{6,487}&\cellcolor{gray!6}{11}&\cellcolor{gray!6}{22}&\cellcolor{gray!6}{0}&\cellcolor{gray!6}{5}&\cellcolor{gray!6}{5}&\cellcolor{gray!6}{3}&\cellcolor{gray!6}{0}&\cellcolor{gray!6}{6,487}&\cellcolor{gray!6}{13}&\cellcolor{gray!6}{20}&\cellcolor{gray!6}{0}&\cellcolor{gray!6}{5}&\cellcolor{gray!6}{5}&\cellcolor{gray!6}{2}\\

\bottomrule

\end{tabular}
\label{table:exceptions}
\vspace{-10pt}
\end{table*}

%% file: sections/FinalStyle/RQs/RQ3.tex
\subsection{\textbf{RQ3: To what extent is the utilization of machine learning helpful in finding the differences between flaky and true failures?}}
\label{rq3}
\input{sections/Tables/Classifier_table}

We evaluate the \classifier using two algorithms (decision tree and Na\"{i}ve Bayes) and two approaches for balancing the dataset.
In terms of balancing the dataset, we use the SMOTE technique if the ratio of one type of failures is less than 10$\%$ of the total number of failures of the other type.
We also consider training on the dataset as it is without balancing.
We use stratified cross-validation and leave one fold for testing purposes.

To further understand the efficacy of machine learning in this context, we looked for a state-of-the-art classifier based on the failure logs. Existing methods to detect flaky failures, like the work of Lampel et al.~\cite{lampel2021life}, do not align with our dataset, which is based on the failure logs. Given this and the discussed features, we considered an alternative baseline approach. We utilized TF-IDF to provide a comparison for our classifier's predictions.
Furthermore, we investigated whether TF-IDF could serve as an alternative method, especially since the features of the \classifier are directly from the syntax of the failure logs without involving dynamic features.

Table~\ref{table:classifier_table} shows the result of using the \classifier and \tfidf in predicting a failure if it is flaky or not.
While we considered two different classification algorithms (Decision Tree and Naive Bayes), we find that decisions trees (without any dataset balancing) performed the best, and present only those results here (complete results are available in our public repository~\cite{failure-log-classifiers-GitHub}).
To ensure that each fold had at least one flaky failure, we include only projects with at least 10 flaky failures in this analysis.
The relative performance of the \classifier and the \tfidf varies as in some projects they have at least 90\% F1 scores with zero \emph{FN} failures while in few projects it is worse than being randomly guessing. The performance of the two classifiers close to each others (78,181 \emph{TP} in the \classifier versus 79,428 in \tfidf \emph{TP}). Both classifiers have less False Positive rates (4,745 in the \classifier and 1,437 in \tfidf) than the rate of using the \syntax (8,587).

Comparing the significant increase in true positives versus the \syntax, we find that \classifier and \tfidf both benefit significantly from the ability to match failures from one test to a different test.
This is because, in our implementation of \syntax, we do \emph{not} remove test-specific lines from the stack trace.
Future work might extend our approaches to abstract these elements out of the stack trace, making matches between tests more likely~\cite{An23JustInTime}.

Comparing \tfidf versus the \classifier, we find that \tfidf performs somewhat better in all cases.
One explanation for this is the presence of line numbers in the failure messages.
As described in Section~\ref{sec:approaches}, we manually filtered this noisy information out from the \syntax and \classifier --- but allowed it to remain for \tfidf to prioritize.
We found including the stacktrace lines numbers added more values as reflecting different stacktraces. 
Classifiers that use more complex features (particularly those that are specific to a project) might be able to outperform \tfidf.
On the other side, the generality of the features that the \classifier could be a reason that, even with high performance in most projects, still not outperform the the \tfidf.

We note two projects where the \classifier performs significantly worse than \tfidf (\wildfly~and \exec).
In the \wildfly, we found all flaky failures with had the exception \emph{RuntimeException} with very low repetitive rate (each test failing at most 7 times) while the same exception often appears in the true failures. 
This project performs well in the \tfidf and even in the \syntax.
The main observation in these failures is that the line numbers in the tests differ, which is not captured from the features we proposed to train the \classifier.
In \exec, we have a similar situation with a frequently-occurring exception, \emph{AssertionFailedError}.

The usability of different machine learning approaches varies based on the specific use case and objectives. 
If the main goal is to maximize the number of true positives (\emph{TP}) without being overly concerned about the rate of false positives, the approach with more \emph{TP} is the better.
In this scenario, the model is more focused on correctly identifying as many flaky failures as possible, even if it means accepting a higher number of false positives.
One of the main advantages of the \classifier is its flexibility in extending the learned features.
The model can be easily augmented with additional static and dynamic features extracted from each failure.
The proposed features shown in Table~\ref{table:Features} are not intended to be a final set for immediate adoption, but serve as a starting point.
By leveraging additional features (particularly those specific to a project), the failure log classifier can potentially enhance its performance identifying flaky failures.

\textbf{Summary.} We found that both the \classifier and \tfidf are able to predict flaky and true failures in most the projects. We found \tfidf is slightly better in terms of the total number of false positives and negatives failures compare to the \classifier result.

%% file: sections/Tables/Classifier_table.tex
\begin{table*}[t]

\caption{Comparison of \classifier and TF-IDF performance on flaky and true failures prediction.\\
\textnormal{For the \classifier and TF-IDF, we show the confusion matrix, precision (P), recall (R), and F1 score of the overall prediction result.
This analysis only includes projects with at least 10 flaky failures (excluding the project ``elasticjob-elastic-job-lite'').
}} 
\label{table:classifier_table}
\vspace{-5pt}
\setlength{\tabcolsep}{3.0pt}
\newcommand{\failureRateWidth}{2.5in}
\newcommand{\failureRateHeight}{4em}
\scriptsize
\centering
    \begin{tabular}{l|rrrr|rrrrrrr|rrrrrrr}
    \toprule
      & \multicolumn{4}{c}{\textbf{Total Flaky Tests and Failures}} & \multicolumn{7}{c}{\textbf{Failure Log Classifier}} & \multicolumn{7}{c}{\textbf{TF-IDF}}\\ 
     
     \textbf{Project}&\textbf{Test}&\textbf{Failures}&\textbf{Flaky}&\textbf{True}&\textbf{TP}&\textbf{FN}&\textbf{FP}&\textbf{TN}&\textbf{P}&\textbf{R}&\textbf{F1}&\textbf{TP}&\textbf{FN}&\textbf{FP}&\textbf{TN}&\textbf{P}&\textbf{R}&\textbf{F1}\\
        \midrule

Alluxio-alluxio&114&49,653&16,858&32,795&16,014&844&1,706&31,089&90\%&94\%&92\%&16,616&242&555&32,240&96\%&98\%&97\%\\
\cellcolor{gray!6}{square-okhttp}&\cellcolor{gray!6}{99}&\cellcolor{gray!6}{60,435}&\cellcolor{gray!6}{26,486}&\cellcolor{gray!6}{33,949}&\cellcolor{gray!6}{26,056}&\cellcolor{gray!6}{430}&\cellcolor{gray!6}{897}&\cellcolor{gray!6}{33,052}&\cellcolor{gray!6}{96\%}&\cellcolor{gray!6}{98\%}&\cellcolor{gray!6}{97\%}&\cellcolor{gray!6}{26,459}&\cellcolor{gray!6}{27}&\cellcolor{gray!6}{104}&\cellcolor{gray!6}{33,845}&\cellcolor{gray!6}{99\%}&\cellcolor{gray!6}{99\%}&\cellcolor{gray!6}{99\%}\\
apache-ambari&51&15,108&4,063&11,045&4,055&8&481&10,564&89\%&99\%&94\%&4,063&0&6&11,039&99\%&100\%&99\%\\
\cellcolor{gray!6}{hector-client-hector}&\cellcolor{gray!6}{33}&\cellcolor{gray!6}{10,132}&\cellcolor{gray!6}{6,529}&\cellcolor{gray!6}{3,603}&\cellcolor{gray!6}{6,529}&\cellcolor{gray!6}{0}&\cellcolor{gray!6}{404}&\cellcolor{gray!6}{3,199}&\cellcolor{gray!6}{94\%}&\cellcolor{gray!6}{100\%}&\cellcolor{gray!6}{96\%}&\cellcolor{gray!6}{6,529}&\cellcolor{gray!6}{0}&\cellcolor{gray!6}{12}&\cellcolor{gray!6}{3,591}&\cellcolor{gray!6}{99\%}&\cellcolor{gray!6}{100\%}&\cellcolor{gray!6}{99\%}\\
activiti-activiti&30&46,300&1,363&44,937&947&416&300&44,637&75\%&69\%&72\%&1,023&340&67&44,870&93\%&75\%&83\%\\
\cellcolor{gray!6}{tootallnate-java-websocket}&\cellcolor{gray!6}{23}&\cellcolor{gray!6}{3,259}&\cellcolor{gray!6}{2,143}&\cellcolor{gray!6}{1,116}&\cellcolor{gray!6}{2,130}&\cellcolor{gray!6}{13}&\cellcolor{gray!6}{436}&\cellcolor{gray!6}{680}&\cellcolor{gray!6}{83\%}&\cellcolor{gray!6}{99\%}&\cellcolor{gray!6}{90\%}&\cellcolor{gray!6}{2,130}&\cellcolor{gray!6}{13}&\cellcolor{gray!6}{437}&\cellcolor{gray!6}{679}&\cellcolor{gray!6}{82\%}&\cellcolor{gray!6}{99\%}&\cellcolor{gray!6}{90\%}\\
apache-httpcore&22&8,375&354&8,021&315&39&108&7,913&74\%&88\%&81\%&314&40&17&8,004&94\%&88\%&91\%\\
\cellcolor{gray!6}{apache-hbase}&\cellcolor{gray!6}{20}&\cellcolor{gray!6}{3,104}&\cellcolor{gray!6}{2,519}&\cellcolor{gray!6}{585}&\cellcolor{gray!6}{2,377}&\cellcolor{gray!6}{142}&\cellcolor{gray!6}{24}&\cellcolor{gray!6}{561}&\cellcolor{gray!6}{99\%}&\cellcolor{gray!6}{94\%}&\cellcolor{gray!6}{96\%}&\cellcolor{gray!6}{2,386}&\cellcolor{gray!6}{133}&\cellcolor{gray!6}{31}&\cellcolor{gray!6}{554}&\cellcolor{gray!6}{98\%}&\cellcolor{gray!6}{94\%}&\cellcolor{gray!6}{96\%}\\
qos-ch-logback&20&3,052&438&2,614&172&266&104&2,510&62\%&39\%&48\%&235&203&37&2,577&86\%&53\%&66\%\\
\cellcolor{gray!6}{kevinsawicki.http-request}&\cellcolor{gray!6}{18}&\cellcolor{gray!6}{3,888}&\cellcolor{gray!6}{3,501}&\cellcolor{gray!6}{387}&\cellcolor{gray!6}{3,498}&\cellcolor{gray!6}{3}&\cellcolor{gray!6}{124}&\cellcolor{gray!6}{263}&\cellcolor{gray!6}{96\%}&\cellcolor{gray!6}{99\%}&\cellcolor{gray!6}{98\%}&\cellcolor{gray!6}{3,498}&\cellcolor{gray!6}{3}&\cellcolor{gray!6}{54}&\cellcolor{gray!6}{333}&\cellcolor{gray!6}{98\%}&\cellcolor{gray!6}{99\%}&\cellcolor{gray!6}{99\%}\\
wildfly-wildfly&18&3,895&48&3,847&0&48&0&3,847&0\%&0\%&0\%&50&0&0&4,364&100\%&100\%&100\%\\
\cellcolor{gray!6}{wro4j-wro4j}&\cellcolor{gray!6}{14}&\cellcolor{gray!6}{11,373}&\cellcolor{gray!6}{10,833}&\cellcolor{gray!6}{540}&\cellcolor{gray!6}{10,833}&\cellcolor{gray!6}{0}&\cellcolor{gray!6}{67}&\cellcolor{gray!6}{473}&\cellcolor{gray!6}{99\%}&\cellcolor{gray!6}{100\%}&\cellcolor{gray!6}{99\%}&\cellcolor{gray!6}{10,833}&\cellcolor{gray!6}{0}&\cellcolor{gray!6}{29}&\cellcolor{gray!6}{511}&\cellcolor{gray!6}{99\%}&\cellcolor{gray!6}{100\%}&\cellcolor{gray!6}{99\%}\\
spring-projects-spring-boot&12&2,164&14&2,150&6&8&0&2,150&100\%&42\%&60\%&9&5&3&2,147&75\%&64\%&69\%\\
\cellcolor{gray!6}{undertow-io-undertow}&\cellcolor{gray!6}{7}&\cellcolor{gray!6}{2,398}&\cellcolor{gray!6}{92}&\cellcolor{gray!6}{2,306}&\cellcolor{gray!6}{3}&\cellcolor{gray!6}{89}&\cellcolor{gray!6}{0}&\cellcolor{gray!6}{2,306}&\cellcolor{gray!6}{100\%}&\cellcolor{gray!6}{3\%}&\cellcolor{gray!6}{6\%}&\cellcolor{gray!6}{4}&\cellcolor{gray!6}{88}&\cellcolor{gray!6}{0}&\cellcolor{gray!6}{2,306}&\cellcolor{gray!6}{100\%}&\cellcolor{gray!6}{4\%}&\cellcolor{gray!6}{8\%}\\
orbit-orbit&7&3,755&2,943&812&2,943&0&69&743&97\%&100\%&98\%&2,943&0&59&753&98\%&100\%&99\%\\
\cellcolor{gray!6}{doanduyhai-Achilles}&\cellcolor{gray!6}{2}&\cellcolor{gray!6}{275}&\cellcolor{gray!6}{121}&\cellcolor{gray!6}{154}&\cellcolor{gray!6}{120}&\cellcolor{gray!6}{1}&\cellcolor{gray!6}{0}&\cellcolor{gray!6}{154}&\cellcolor{gray!6}{100\%}&\cellcolor{gray!6}{99\%}&\cellcolor{gray!6}{99\%}&\cellcolor{gray!6}{120}&\cellcolor{gray!6}{1}&\cellcolor{gray!6}{0}&\cellcolor{gray!6}{154}&\cellcolor{gray!6}{100\%}&\cellcolor{gray!6}{99\%}&\cellcolor{gray!6}{99\%}\\
joel-costigliola-assertj-core&1&992&974&18&974&0&1&17&99\%&100\%&99\%&974&0&0&18&100\%&100\%&100\%\\
\cellcolor{gray!6}{jknack-handlebars.java}&\cellcolor{gray!6}{1}&\cellcolor{gray!6}{558}&\cellcolor{gray!6}{411}&\cellcolor{gray!6}{147}&\cellcolor{gray!6}{411}&\cellcolor{gray!6}{0}&\cellcolor{gray!6}{16}&\cellcolor{gray!6}{131}&\cellcolor{gray!6}{96\%}&\cellcolor{gray!6}{100\%}&\cellcolor{gray!6}{98\%}&\cellcolor{gray!6}{411}&\cellcolor{gray!6}{0}&\cellcolor{gray!6}{16}&\cellcolor{gray!6}{131}&\cellcolor{gray!6}{96\%}&\cellcolor{gray!6}{100\%}&\cellcolor{gray!6}{98\%}\\
ninjaframework-ninja&1&596&476&120&476&0&8&112&98\%&100\%&99\%&476&0&8&112&98\%&100\%&99\%\\
\cellcolor{gray!6}{zxing-zxing}&\cellcolor{gray!6}{1}&\cellcolor{gray!6}{398}&\cellcolor{gray!6}{322}&\cellcolor{gray!6}{76}&\cellcolor{gray!6}{322}&\cellcolor{gray!6}{0}&\cellcolor{gray!6}{0}&\cellcolor{gray!6}{76}&\cellcolor{gray!6}{100\%}&\cellcolor{gray!6}{100\%}&\cellcolor{gray!6}{100\%}&\cellcolor{gray!6}{322}&\cellcolor{gray!6}{0}&\cellcolor{gray!6}{0}&\cellcolor{gray!6}{76}&\cellcolor{gray!6}{100\%}&\cellcolor{gray!6}{100\%}&\cellcolor{gray!6}{100\%}\\
apache-commons-exec&1&92&33&59&0&33&0&59&0\%&0\%&0\%&33&0&2&57&94\%&100\%&97\%\\
\midrule
\cellcolor{gray!6}{21 Total Projects}&\cellcolor{gray!6}{495}&\cellcolor{gray!6}{229,802}&\cellcolor{gray!6}{80,521}&\cellcolor{gray!6}{149,281}&\cellcolor{gray!6}{78,181}&\cellcolor{gray!6}{2,340}&\cellcolor{gray!6}{4,745}&\cellcolor{gray!6}{144,536}&\cellcolor{gray!6}{}&\cellcolor{gray!6}{}&\cellcolor{gray!6}{}&\cellcolor{gray!6}{79,428}&\cellcolor{gray!6}{1,095}&\cellcolor{gray!6}{1,437}&\cellcolor{gray!6}{148,361}&\cellcolor{gray!6}{}&\cellcolor{gray!6}{}&\cellcolor{gray!6}{}\\

\bottomrule

\end{tabular}
\vspace{-10pt}
\end{table*}

%% file: sections/FinalStyle/threats.tex
\section{Threats to Validity}
\label{threats}

We construct an experiment to evaluate the efficacy of approaches that determines whether or not a test failure is flaky based on matching that failure against previously witnessed failures.
This methodology accurately represents cases where developers already have a historical set of failures, and are evaluating new failures as they arrive.
However: developers may be more interested in other usecases, which we did not evaluate.
In a real-world scenario, flaky failures could be a collected by merging diverse failure logs collected in CI, in different runtime environments, different timeframes, and even different code revisions under testing.
Based on data availability, our evaluation methodology collects failures from a single code revision --- we hypothesize that our key findings regarding the repetitive nature of flaky failures would generalize, but leave such a study for future work.
Our evaluation methodology also assumes that a significant proportion of failures have been labeled as true or flaky failures using existing methodologies.
However, developers may not have all the failures being labeled.

We use failures of tests detected during mutation analysis as a stand-in for true regression failures, as we were unable to prepare a dataset containing both flaky and true regression failures for multiple projects.
However, studies have repeatedly demonstrated that mutants are a valid substitute for real faults in many software testing contexts~\cite{just2014mutants}.
Moreover, our results are complemented by existing case studies of individual projects like Google Chromium~\cite{haben2023importance} and SAP HANA~\cite{An23JustInTime}.

Our dataset of tests is drawn from existing work~\cite{alshammari2021flakeflagger}, and has been re-used in other recent works as well~\cite{9978221,DellAnna22Evaluating,Pontillo22Static}.
This dataset includes projects from different domains, but all projects are implemented in Java.
The overall performance of these failure matching approaches may vary between different languages and testing frameworks.
We leave an extension of this study to other languages to future work.
Based on our results, it is clear that the performance of flaky failure prediction approaches will vary across projects.
While our experiment demonstrates a full range of performance of these approaches (from near-perfect performance to extremely poor performance), it is difficult to extrapolate what the ``average'' case would be.
We take care to avoid drawing such conclusions, and instead aim to identify patterns in projects, tests, and failures that may impact the performance of these approaches.

%% file: sections/FinalStyle/discussion.tex
\section{Discussion}
\label{discussion}

By examining the exception types and frequency, we concluded that the causes of flaky failures in our dataset range from code-related issues to environmental factors.
Experienced developers can sometimes easily identify if a failure is flaky just by examining its log~\cite{gradlePreventingFlaky}, indicating that certain failures are readily detectable as flaky.
Our examination of exception types confirmed that detecting these environmental-related failures can be easy.

However, our evaluation also revealed a category of failures that are \emph{hard} to classify as flaky or not --- in particular, failures with an \emph{AssertionFailedError} or \emph{NullPointerException}.
Failures with these kinds of exceptions appeared to be fairly low in information: the failure does not provide enough information to determine if it is caused by flakiness or not.
One approach to improve flakiness detection for these kinds of failures might be to enhance the tests or system under test.
By providing richer logging information about the symptoms that led to the failure, matching-based approaches may be able to discriminate better between flaky and non-flaky failures.
Future work might study these failures further.

Using machine learning in addressing this problem may offer another solution.
In our study, we use the decision tree algorithm due to its ease of implementation, particularly in datasets with a minimal number of features, with the majority being boolean features.
However, it is worth considering other supervised learning algorithms if the user of this approach intends to expand the feature list to include additional details. 
Mining additional features is likely to improve prediction performance, although we expect it would be challenging to uniformly improve performance across many projects.

%% file: sections/FinalStyle/relatedWork.tex
\section{Related Work}

\textbf{Detecting Flaky Failures.} Rerunning failing tests has been a de-facto approach for developers to determine if a test failure is caused by flakiness, or is a true failure~\cite{Micco17State,MavenRerunFailing}.
Bell et al. studied the efficacy of different rerunning strategies, finding that simply re-running failing tests immediately upon observing a failure is often ineffective at confirming that a failure is flaky \cite{bell2018deflaker}.
They proposed \emph{DeFlaker}, an alternative strategy for determining if a failure is flaky or not by intersecting the line coverage of each failing test with the set of lines that changed since the last test suite execution.
If a test fails but does not cover any changed lines, it can be confirmed as flaky without being re-run.
While this approach may be elegant, it can be challenging to apply in practice, as it requires developers to use a specialized code coverage instrumentation agent.
Moreover, while Bell et al. show that the approach is faster than typical code coverage agents, it still imposes a runtime overhead of up to 12\%.
In contrast, in this article, we evaluate approaches for determining whether a test failure is caused by flakiness \emph{without} requiring any code instrumentation, and imposing no overhead on test execution.

Concurrent to our work, two other research teams have been working on this same important challenge.
Haben et al. empirically demonstrate the importance of determining precisely which failures of flaky tests are true by showing the danger of assuming that all failures of flaky tests are to be ignored~\cite{haben2023importance}.
This work relies on training classifiers from the \emph{code} of flaky tests, while we rely on approaches that utilize the \emph{failures} of flaky tests.
More similar to our approach, An et al. use abstracted information from error messages and stack traces to determine which test failures are flaky in the SAP HANA database~\cite{An23JustInTime}.
An et al. abstract failure symptoms using techniques similar to those that we use, for example, removing line numbers and test entry points from stack traces. 
On SAP HANA, they report a precision of 96\% and recall of 76\% in detecting flaky failures, results that are comparable to those some of the open-source projects that we evaluated.
We believe that these two lines of work are quite complementary, with our work providing a replicable open-source dataset of failures on multiple projects, and An et al's work providing a deep case study of how to successfully apply flaky failure detection in production at a large software company.

Flaky tests often surface when test suite are run in continuous integration (CI) platforms.
With CI, each revision of the system under test is automatically built and tested.
Flaky tests can be a nuisance by resulting in builds appearing to have ``failed'' when they would in fact have passed, if not for flakiness.
While our approach aims to determine which failures of which tests are caused by flakiness, other approaches have studied this problem at the level of entire CI builds.
Lampel et al. study intermittent job failures in the Mozilla CI platform, training models based on various telemetry (including runtime, CPU load and OS version) to determine which failed builds are flaky.
In contrast, Olewick et al. utilize a bag-of-words model, extracting a vocabulary from each CI build log in order to determine which builds failed due to flakiness~\cite{Olewicki22Towards}.
An advantage to these approaches is that they are language and platform-agnostic, requiring \emph{no} knowledge of the structure of log files.
However, a corresponding disadvantage is that they may achieve lower predictive performance: An et al. conduct an ablation study examining the importance of failure abstraction, finding performance to drop by 50\%~\cite{An23JustInTime}.

\textbf{Detecting Flaky Tests.}
A related line of research aims to determine not which test \emph{failures} are flaky, but which tests \emph{could} fail due to flakiness.
If tools could inform developers that a test is flaky immediately as the test is being created, then perhaps flaky tests could be avoided all together.
One class of approaches aim to detect flaky tests that are flaky due to a particular root cause.
For example: \emph{NonDex} proactively detects flaky tests that rely on non-deterministic behavior (such as the order of iteration of an unsorted collection)~\cite{Gyori16NonDex}.
\emph{iDFlakies} detects order-dependent flaky tests --- those that have flaky failures when executed in different orders~\cite{lam2019idflakies}.

Given the goal of broadly detecting which tests might be flaky (not tied to a specific root cause), a baseline approach is to re-run a test suite many times.
Alshammari et al. examined a dataset of 26 open-source Java projects, re-running each test suite 10,000 times to identify which tests could be flaky~\cite{alshammari2021flakeflagger}.
They proposed \emph{FlakeFlagger}, a machine learning-based approach to determine which tests might be flaky using a set of features collected while tests run.
Other flaky test prediction approaches rely on the \emph{vocabulary} of test methods, with the insight that tests that perform tasks similar to flaky tests are also flaky\cite{Pinto20WhatIs,Verdecchia21Know,9978221,DellAnna22Evaluating,Pontillo22Static}.
Parry et al. extend this body of work by demonstrating how to efficiently combine machine learning-based flaky test detection with test reruns~\cite{Parry23Empirically}.
Our goal is not to detect which \emph{tests} could be flaky, but rather, which \emph{failures} are flaky, assuming that a developer has previously identified a set of flaky test failures.

\textbf{Test Failure Clustering.} We evaluate a \syntax approach for determining if a test failure is flaky or not.
This approach is inspired by work in a closely related area of failure de-duplication.
Classic approaches in this field aim to automatically group multiple test failures together by a shared root cause using stack traces and/or failure logs \cite{10.5555/1855895.1855896,6227111,6498456,Jiang17WhatCauses,Podgurski03Automated}.
However, applying this approach to the problem of failure flakiness detection is relatively under-studied.
Lam et al. describe a Microsoft-internal tool that suppresses failures of known flaky tests, and suggests that the error message is used as part of that process~\cite{Lam20Study}.
However, they provide no evaluation of how often this approach incorrectly suppresses a test failure that should be investigated as a true failure.
We advance this field of study by providing a detailed analysis of the efficacy of test failure clustering for the purposes of labeling test failures as flaky or not.
We make our entire dataset and pipeline publicly available along with this article, in order to allow other researchers to study more advanced test failure clustering applications for this use-case.

%% file: sections/FinalStyle/conclusion.tex
\section{Conclusion}
\label{conclusion}
Using a novel methodology, we constructed a ground-truth dataset of 149,909 true (non-flaky) failures and 80,530 flaky failures from 22 open-source Java projects.
Our analysis shows that, even for the same test, there can be multiple flaky failure symptoms, but that a small set of failing stack traces often reoccur.
This finding provides strong evidence that heuristic-based approaches that determine whether a failure is caused by flakiness or a true defect can be effective. 
Our evaluation of three heuristic approaches for determining whether a test failure is flaky or not showed that performance can vary widely between projects, and that TF-IDF is the best approach overall.

Our results show that some projects may be able to adopt these approaches immediately with no (or almost no) false positives.
In other projects where the failure logs lack informative details, like failures with the presence of assertions statements, the approaches may not be effective.
Increasing the amount of information in test failure logs can greatly improve the performance of automated approaches for de-duplicatign failures, and can further help with manual analysis.

Given the variability between different projects and inherent non-determinism of flaky tests, it will be challenging to create general-purpose solutions for determining whether a failure is flaky or not.
Instrumentation-based approaches (such as DeFlaker~\cite{Bell18Deflaker}) can help add important context to otherwise low-information test failures, but deploying them in production can bring challenges.
Future work might continue to study the application of these approaches as case studies in single projects (e.g. recent work studying Google Chromium~\cite{haben2023importance} and SAP HANA~\cite{An23JustInTime}).
We make our entire dataset and experiments available under an open-source license to enable and encourage future research in this important topic.

\section{Data Availability Statement}

All analysis notebooks and dataset related to this work are publicly available at~\cite{failure-log-classifiers-GitHub} ~\cite{Failure-Logs-Dataset}.

\section{Acknowledgement}

We thank Kevin Moran and Wing Lam for their valuable discussions related to this work. 

%% file: main.bbl
\begin{thebibliography}{10}
\providecommand{\url}[1]{#1}
\csname url@samestyle\endcsname
\providecommand{\newblock}{\relax}
\providecommand{\bibinfo}[2]{#2}
\providecommand{\BIBentrySTDinterwordspacing}{\spaceskip=0pt\relax}
\providecommand{\BIBentryALTinterwordstretchfactor}{4}
\providecommand{\BIBentryALTinterwordspacing}{\spaceskip=\fontdimen2\font plus
\BIBentryALTinterwordstretchfactor\fontdimen3\font minus \fontdimen4\font\relax}
\providecommand{\BIBforeignlanguage}[2]{{%
\expandafter\ifx\csname l@#1\endcsname\relax
\typeout{** WARNING: IEEEtran.bst: No hyphenation pattern has been}%
\typeout{** loaded for the language `#1'. Using the pattern for}%
\typeout{** the default language instead.}%
\else
\language=\csname l@#1\endcsname
\fi
#2}}
\providecommand{\BIBdecl}{\relax}
\BIBdecl

\bibitem{Parry21Survey}
\BIBentryALTinterwordspacing
O.~Parry, G.~M. Kapfhammer, M.~Hilton, and P.~McMinn, ``A survey of flaky tests,'' \emph{ACM Trans. Softw. Eng. Methodol.}, vol.~31, no.~1, oct 2021. [Online]. Available: \url{https://doi.org/10.1145/3476105}
\BIBentrySTDinterwordspacing

\bibitem{Rahman18Impact}
\BIBentryALTinterwordspacing
M.~T. Rahman and P.~C. Rigby, ``The impact of failing, flaky, and high failure tests on the number of crash reports associated with firefox builds,'' in \emph{Proceedings of the 2018 26th ACM Joint Meeting on European Software Engineering Conference and Symposium on the Foundations of Software Engineering}, ser. ESEC/FSE 2018.\hskip 1em plus 0.5em minus 0.4em\relax New York, NY, USA: Association for Computing Machinery, 2018, p. 857–862. [Online]. Available: \url{https://doi.org/10.1145/3236024.3275529}
\BIBentrySTDinterwordspacing

\bibitem{haben2023importance}
G.~Haben, S.~Habchi, M.~Papadakis, M.~Cordy, and Y.~L. Traon, ``The importance of discerning flaky from fault-triggering test failures: A case study on the chromium ci,'' \emph{arXiv preprint arXiv:2302.10594}, 2023.

\bibitem{Gruber22Survey}
\BIBentryALTinterwordspacing
M.~Gruber and G.~Fraser, ``A survey on how test flakiness affects developers and what support they need to address it,'' in \emph{2022 IEEE Conference on Software Testing, Verification and Validation (ICST)}.\hskip 1em plus 0.5em minus 0.4em\relax Los Alamitos, CA, USA: IEEE Computer Society, apr 2022, pp. 82--92. [Online]. Available: \url{https://doi.ieeecomputersociety.org/10.1109/ICST53961.2022.00020}
\BIBentrySTDinterwordspacing

\bibitem{Eck19Understanding}
\BIBentryALTinterwordspacing
M.~Eck, F.~Palomba, M.~Castelluccio, and A.~Bacchelli, ``Understanding flaky tests: The developer’s perspective,'' in \emph{Proceedings of the 2019 27th ACM Joint Meeting on European Software Engineering Conference and Symposium on the Foundations of Software Engineering}, ser. ESEC/FSE 2019.\hskip 1em plus 0.5em minus 0.4em\relax New York, NY, USA: Association for Computing Machinery, 2019, p. 830–840. [Online]. Available: \url{https://doi.org/10.1145/3338906.3338945}
\BIBentrySTDinterwordspacing

\bibitem{habchi2022qualitative}
S.~Habchi, G.~Haben, M.~Papadakis, M.~Cordy, and Y.~Le~Traon, ``A qualitative study on the sources, impacts, and mitigation strategies of flaky tests,'' in \emph{2022 IEEE Conference on Software Testing, Verification and Validation (ICST)}.\hskip 1em plus 0.5em minus 0.4em\relax IEEE, 2022, pp. 244--255.

\bibitem{Lam20Understanding}
W.~Lam, S.~Winter, A.~Astorga, V.~Stodden, and D.~Marinov, ``Understanding reproducibility and characteristics of flaky tests through test reruns in java projects,'' in \emph{2020 IEEE 31st International Symposium on Software Reliability Engineering (ISSRE)}, 2020, pp. 403--413.

\bibitem{Bell18Deflaker}
\BIBentryALTinterwordspacing
J.~Bell, O.~Legunsen, M.~Hilton, L.~Eloussi, T.~Yung, and D.~Marinov, ``Deflaker: Automatically detecting flaky tests,'' in \emph{Proceedings of the 40th International Conference on Software Engineering}, ser. ICSE '18.\hskip 1em plus 0.5em minus 0.4em\relax New York, NY, USA: Association for Computing Machinery, 2018, p. 433–444. [Online]. Available: \url{https://doi.org/10.1145/3180155.3180164}
\BIBentrySTDinterwordspacing

\bibitem{Micco17State}
J.~Micco, ``The state of continuous integration testing @ google,'' \url{https://storage.googleapis.com/pub-tools-public-publication-data/pdf/45880.pdf}.

\bibitem{alshammari2021flakeflagger}
A.~Alshammari, C.~Morris, M.~Hilton, and J.~Bell, ``Flakeflagger: Predicting flakiness without rerunning tests,'' in \emph{2021 IEEE/ACM 43rd International Conference on Software Engineering (ICSE)}.\hskip 1em plus 0.5em minus 0.4em\relax IEEE, 2021, pp. 1572--1584.

\bibitem{bell2018deflaker}
J.~Bell, O.~Legunsen, M.~Hilton, L.~Eloussi, T.~Yung, and D.~Marinov, ``Deflaker: Automatically detecting flaky tests,'' in \emph{2018 IEEE/ACM 40th International Conference on Software Engineering (ICSE)}.\hskip 1em plus 0.5em minus 0.4em\relax IEEE, 2018, pp. 433--444.

\bibitem{lam2019idflakies}
W.~Lam, R.~Oei, A.~Shi, D.~Marinov, and T.~Xie, ``idflakies: A framework for detecting and partially classifying flaky tests,'' in \emph{2019 12th ieee conference on software testing, validation and verification (icst)}.\hskip 1em plus 0.5em minus 0.4em\relax IEEE, 2019, pp. 312--322.

\bibitem{Pontillo22Static}
\BIBentryALTinterwordspacing
V.~Pontillo, F.~Palomba, and F.~Ferrucci, ``Static test flakiness prediction: How far can we go?'' \emph{Empirical Softw. Engg.}, vol.~27, no.~7, dec 2022. [Online]. Available: \url{https://doi.org/10.1007/s10664-022-10227-1}
\BIBentrySTDinterwordspacing

\bibitem{Verdecchia21Know}
R.~Verdecchia, E.~Cruciani, B.~Miranda, and A.~Bertolino, ``Know you neighbor: Fast static prediction of test flakiness,'' \emph{IEEE Access}, vol.~9, pp. 76\,119--76\,134, 2021.

\bibitem{Parry23Empirically}
\BIBentryALTinterwordspacing
O.~Parry, G.~M. Kapfhammer, M.~Hilton, and P.~McMinn, ``Empirically evaluating flaky test detection techniques combining test case rerunning and machine learning models,'' \emph{Empirical Softw. Engg.}, vol.~28, no.~3, apr 2023. [Online]. Available: \url{https://doi.org/10.1007/s10664-023-10307-w}
\BIBentrySTDinterwordspacing

\bibitem{Lam20Study}
\BIBentryALTinterwordspacing
W.~Lam, K.~Mu\c{s}lu, H.~Sajnani, and S.~Thummalapenta, ``A study on the lifecycle of flaky tests,'' in \emph{Proceedings of the ACM/IEEE 42nd International Conference on Software Engineering}, ser. ICSE '20.\hskip 1em plus 0.5em minus 0.4em\relax New York, NY, USA: Association for Computing Machinery, 2020, p. 1471–1482. [Online]. Available: \url{https://doi.org/10.1145/3377811.3381749}
\BIBentrySTDinterwordspacing

\bibitem{Gyori16NonDex}
\BIBentryALTinterwordspacing
A.~Gyori, B.~Lambeth, A.~Shi, O.~Legunsen, and D.~Marinov, ``Nondex: A tool for detecting and debugging wrong assumptions on java api specifications,'' in \emph{Proceedings of the 2016 24th ACM SIGSOFT International Symposium on Foundations of Software Engineering}, ser. FSE 2016.\hskip 1em plus 0.5em minus 0.4em\relax New York, NY, USA: Association for Computing Machinery, 2016, p. 993–997. [Online]. Available: \url{https://doi.org/10.1145/2950290.2983932}
\BIBentrySTDinterwordspacing

\bibitem{An23JustInTime}
G.~An, J.~Yoon, T.~Bach, J.~Hong, and S.~Yoo, ``Just-in-time flaky test detection via abstracted failure symptom matching,'' 2023.

\bibitem{Podgurski03Automated}
A.~Podgurski, D.~Leon, P.~Francis, W.~Masri, M.~Minch, J.~Sun, and B.~Wang, ``Automated support for classifying software failure reports,'' in \emph{Proceedings of the 25th International Conference on Software Engineering}, ser. ICSE '03.\hskip 1em plus 0.5em minus 0.4em\relax USA: IEEE Computer Society, 2003, p. 465–475.

\bibitem{Jiang17WhatCauses}
\BIBentryALTinterwordspacing
H.~Jiang, X.~Li, Z.~Yang, and J.~Xuan, ``What causes my test alarm? automatic cause analysis for test alarms in system and integration testing,'' in \emph{Proceedings of the 39th International Conference on Software Engineering}, ser. ICSE '17.\hskip 1em plus 0.5em minus 0.4em\relax IEEE Press, 2017, p. 712–723. [Online]. Available: \url{https://doi.org/10.1109/ICSE.2017.71}
\BIBentrySTDinterwordspacing

\bibitem{gradlePreventingFlaky}
D.~Welter, ``{P}reventing {F}laky {T}ests from {R}uining your {T}est {S}uite --- gradle.com,'' \url{https://gradle.com/blog/prevent-flaky-tests/}, [Accessed 29-May-2023].

\bibitem{lampel2021life}
J.~Lampel, S.~Just, S.~Apel, and A.~Zeller, ``When life gives you oranges: detecting and diagnosing intermittent job failures at mozilla,'' in \emph{Proceedings of the 29th ACM Joint Meeting on European Software Engineering Conference and Symposium on the Foundations of Software Engineering}, 2021, pp. 1381--1392.

\bibitem{DT}
A.~Navada, A.~N. Ansari, S.~Patil, and B.~A. Sonkamble, ``Overview of use of decision tree algorithms in machine learning,'' in \emph{2011 IEEE control and system graduate research colloquium}.\hskip 1em plus 0.5em minus 0.4em\relax IEEE, 2011, pp. 37--42.

\bibitem{tfidf}
J.~Ramos \emph{et~al.}, ``Using tf-idf to determine word relevance in document queries,'' in \emph{Proceedings of the first instructional conference on machine learning}, vol. 242, no.~1.\hskip 1em plus 0.5em minus 0.4em\relax Citeseer, 2003, pp. 29--48.

\bibitem{tfidf1}
L.-P. Jing, H.-K. Huang, and H.-B. Shi, ``Improved feature selection approach tfidf in text mining,'' in \emph{Proceedings. International Conference on Machine Learning and Cybernetics}, vol.~2.\hskip 1em plus 0.5em minus 0.4em\relax IEEE, 2002, pp. 944--946.

\bibitem{just2014defects4j}
R.~Just, D.~Jalali, and M.~D. Ernst, ``Defects4j: A database of existing faults to enable controlled testing studies for java programs,'' in \emph{Proceedings of the 2014 international symposium on software testing and analysis}, 2014, pp. 437--440.

\bibitem{saha2018bugs}
R.~K. Saha, Y.~Lyu, W.~Lam, H.~Yoshida, and M.~R. Prasad, ``Bugs. jar: A large-scale, diverse dataset of real-world java bugs,'' in \emph{Proceedings of the 15th international conference on mining software repositories}, 2018, pp. 10--13.

\bibitem{tomassi2019bugswarm}
D.~A. Tomassi, N.~Dmeiri, Y.~Wang, A.~Bhowmick, Y.-C. Liu, P.~T. Devanbu, B.~Vasilescu, and C.~Rubio-Gonz{\'a}lez, ``Bugswarm: Mining and continuously growing a dataset of reproducible failures and fixes,'' in \emph{2019 IEEE/ACM 41st International Conference on Software Engineering (ICSE)}.\hskip 1em plus 0.5em minus 0.4em\relax IEEE, 2019, pp. 339--349.

\bibitem{bears}
\BIBentryALTinterwordspacing
F.~Madeiral, S.~Urli, M.~Maia, and M.~Monperrus, ``Bears: An extensible java bug benchmark for automatic program repair studies,'' in \emph{2019 IEEE 26th International Conference on Software Analysis, Evolution and Reengineering (SANER)}.\hskip 1em plus 0.5em minus 0.4em\relax Los Alamitos, CA, USA: IEEE Computer Society, feb 2019, pp. 468--478. [Online]. Available: \url{https://doi.ieeecomputersociety.org/10.1109/SANER.2019.8667991}
\BIBentrySTDinterwordspacing

\bibitem{just2014mutants}
R.~Just, D.~Jalali, L.~Inozemtseva, M.~D. Ernst, R.~Holmes, and G.~Fraser, ``Are mutants a valid substitute for real faults in software testing?'' in \emph{Proceedings of the 22nd ACM SIGSOFT International Symposium on Foundations of Software Engineering}, 2014, pp. 654--665.

\bibitem{shi2019mitigating}
\BIBentryALTinterwordspacing
A.~Shi, J.~Bell, and D.~Marinov, ``Mitigating the effects of flaky tests on mutation testing,'' in \emph{Proceedings of the 28th ACM SIGSOFT International Symposium on Software Testing and Analysis}, ser. ISSTA 2019.\hskip 1em plus 0.5em minus 0.4em\relax New York, NY, USA: Association for Computing Machinery, 2019, pp. 112--122. [Online]. Available: \url{https://doi.org/10.1145/3293882.3330568}
\BIBentrySTDinterwordspacing

\bibitem{coles2016pit}
H.~Coles, T.~Laurent, C.~Henard, M.~Papadakis, and A.~Ventresque, ``Pit: a practical mutation testing tool for java,'' in \emph{Proceedings of the 25th international symposium on software testing and analysis}, 2016, pp. 449--452.

\bibitem{SMOTE}
A.~Fern{\'a}ndez, S.~Garcia, F.~Herrera, and N.~V. Chawla, ``Smote for learning from imbalanced data: progress and challenges, marking the 15-year anniversary,'' \emph{Journal of artificial intelligence research}, vol.~61, pp. 863--905, 2018.

\bibitem{crossValidation}
D.~Berrar \emph{et~al.}, ``Cross-validation.'' 2019.

\bibitem{failure-log-classifiers-GitHub}
A.~Alshammari, P.~Ammann, M.~Hilton, and J.~Bell, ``Failure log classifiers,'' \url{https://github.com/AlshammariA/FailureLogClassifiers}, 2024.

\bibitem{9978221}
Y.~Qin, S.~Wang, K.~Liu, B.~Lin, H.~Wu, L.~Li, X.~Mao, and T.~F. Bissyande, ``Peeler: Learning to effectively predict flakiness without running tests,'' in \emph{2022 IEEE International Conference on Software Maintenance and Evolution (ICSME)}, 2022, pp. 257--268.

\bibitem{DellAnna22Evaluating}
\BIBentryALTinterwordspacing
D.~Dell'Anna, F.~B. Aydemir, and F.~Dalpiaz, ``Evaluating classifiers in se research: the ecser pipeline and two replication studies,'' \emph{Empirical Software Engineering}, vol.~28, no.~1, p.~3, Nov 2022. [Online]. Available: \url{https://doi.org/10.1007/s10664-022-10243-1}
\BIBentrySTDinterwordspacing

\bibitem{MavenRerunFailing}
\BIBentryALTinterwordspacing
``{Maven Surefire plugin - Rerun failing tests},'' \url{https://maven.apache.org/surefire/maven-surefire-plugin/examples/rerun-failing-tests.html}, 2023. [Online]. Available: \url{https://maven.apache.org/surefire/maven-surefire-plugin/examples/rerun-failing-tests.html}
\BIBentrySTDinterwordspacing

\bibitem{Olewicki22Towards}
D.~Olewicki, M.~Nayrolles, and B.~Adams, ``Towards language-independent brown build detection,'' in \emph{2022 IEEE/ACM 44th International Conference on Software Engineering (ICSE)}, 2022, pp. 2177--2188.

\bibitem{Pinto20WhatIs}
\BIBentryALTinterwordspacing
G.~Pinto, B.~Miranda, S.~Dissanayake, M.~d'Amorim, C.~Treude, and A.~Bertolino, ``What is the vocabulary of flaky tests?'' in \emph{Proceedings of the 17th International Conference on Mining Software Repositories}, ser. MSR '20.\hskip 1em plus 0.5em minus 0.4em\relax New York, NY, USA: Association for Computing Machinery, 2020, p. 492–502. [Online]. Available: \url{https://doi.org/10.1145/3379597.3387482}
\BIBentrySTDinterwordspacing

\bibitem{10.5555/1855895.1855896}
K.~Bartz, J.~W. Stokes, J.~C. Platt, R.~Kivett, D.~Grant, S.~Calinoiu, and G.~Loihle, ``Finding similar failures using callstack similarity,'' in \emph{Proceedings of the Third Conference on Tackling Computer Systems Problems with Machine Learning Techniques}, ser. SysML'08.\hskip 1em plus 0.5em minus 0.4em\relax USA: USENIX Association, 2008, p.~1.

\bibitem{6227111}
Y.~Dang, R.~Wu, H.~Zhang, D.~Zhang, and P.~Nobel, ``Rebucket: A method for clustering duplicate crash reports based on call stack similarity,'' in \emph{2012 34th International Conference on Software Engineering (ICSE)}, 2012, pp. 1084--1093.

\bibitem{6498456}
J.~Lerch and M.~Mezini, ``Finding duplicates of your yet unwritten bug report,'' in \emph{2013 17th European Conference on Software Maintenance and Reengineering}, 2013, pp. 69--78.

\bibitem{Failure-Logs-Dataset}
\BIBentryALTinterwordspacing
A.~Alshammari, P.~Ammann, M.~Hilton, and J.~Bell, ``{Flaky and True Failures Logs to Accompany "230,439 Test Failures Later: An Empirical Evaluation of Flaky Failure Classifiers"},'' Jan. 2024. [Online]. Available: \url{https://doi.org/10.5281/zenodo.10531160}
\BIBentrySTDinterwordspacing

\end{thebibliography}
